\documentclass[12pt]{iopart}

\usepackage{amsgen}
\usepackage{amssymb}
\usepackage{setstack}
\usepackage{bbold}
\usepackage{mathrsfs}
\usepackage{graphicx}
\usepackage{epsfig}
\usepackage{epstopdf}
\usepackage{cite}
\usepackage{color}
\usepackage{rotating}
\usepackage{hyperref}
\usepackage{tikz}

\usetikzlibrary{arrows.meta,arrows}

\newcommand{\eqref}[1]{(\ref{#1})}
\newcommand{\h}[1]{\hat{#1}}

\begin{document}

\title[The Momentum Constraint Equation in PPNC]{The momentum constraint equation in Parameterised Post-Newtonian Cosmology}

\author{Theodore Anton$^1$ and Timothy Clifton$^2$}
\address{School of Physics \& Astronomy, Queen Mary University of London, UK.}
\ead{$^1$t.j.anton@qmul.ac.uk, $^2$t.clifton@qmul.ac.uk}

\begin{abstract}

We derive a theory-independent version of the momentum constraint equation for use in cosmology, as a part of the Parameterised Post-Newtonian Cosmology (PPNC) framework. Our equations are constructed by adapting the corresponding quantities from formalisms constructed for testing and constraining gravity in isolated astrophysical systems, thereby extending the domain of applicability of these approaches up to cosmological scales. Our parameterised equations include both scalar and divergenceless-vector gravitational potentials, and can be applied to both conservative and non-conservative theories of gravity. They can also be used to describe the gravitational fields of both non-linear structures and super-horizon perturbations.
We apply the parameterised equations we propose to quintessence models of dark energy, as well as scalar-tensor and vector-tensor theories of gravity. We find them to work well in each case. Our equations are highly compact, and are intended to be useful for constraining gravity in a theory-independent fashion in cosmology.

\end{abstract}

\section{Introduction}\label{sec:introduction}

General Relativity (GR) is being tested to ever higher precision in a variety of astrophysical and cosmological settings. Cosmological surveys allow gravity to be tested on large scales in the universe \cite{Ishak_2018}, gravitational wave observations are directly probing the strong-field  \cite{GW150914}, and Solar System observations are probing the gravitational field within our own vicinity \cite{Will_1993, Poisson_2014}. Motivation for this work comes from specific perceived shortcomings of GR, such as its apparent need for Dark Energy in order to account for cosmological observations, and the ultimate need to construct a viable quantum theory of gravity, as well as from the scientific requirement for us to experimentally test our hypotheses.

One can, of course, pick specific theories of gravity and calculate predictions for the gravitational phenomena that should result. These can then be compared to observations, and inferences made regarding the viability (or otherwise) of the particular theory in question. While straightforward, this approach to model testing can only ever tell us about the goodness-of-fit of the predictions of any specific theory to the data, and may not always allow us to explore the infinite-dimensional theory space of possibilities. A more useful approach in this regard is to construct theory-independent frameworks that can be used to interpret observational data. The idea behind such approaches is to explore all (or at least some) of the possible deviations from Einstein's theory, without having to specify a particular alternative theory in mind.

Theory-independent approaches serve as a useful halfway house between observational data and specific theories. They provide observers with a set of parameters that can be constrained by their data, and they provide theorists with a set of constraints on possible deviations from GR. This allows the viability of newly proposed theories to be readily evaluated, and indeed provides theoretical physicists with a set of guidelines as to the types of generalisations of Einstein's theory that might profitably be considered. It also gives all of us some concrete idea about the phenomenological consequences of particular types of deviations from Einstein's theory, by establishing which sets of observations can probe which specific types of deviations from GR.

The Parameterised Post-Newtonian (PPN) formalism is the most successful of these theory-independent frameworks \cite{Will_2014}. Within the Solar System, and in extrasolar systems, this formalism has been used to interpret a wide array of gravitational phenomena \cite{Bertotti_2003, Biswas_2008, Gravity_Probe_B, Weisberg_1984}, and has been used to placed strong constraints on possible deviations from GR. Some of the reasons for this success are the simplicity of the PPN formalism, as well as its relative insensitivity to the finer details of the gravitational theories: it expresses the entire phenomenology of relativistic gravity in terms of a small set of measurable parameters, constraints on which apply to a wide array of possible deviations from GR. It assumes only basic properties of gravity in order to do this, in particular the Einstein equivalence principle and the conservation of stress-energy.

In this paper, we continue the development of a framework which imports this highly successful approach to understanding gravitational phenomena into a cosmological context. We call the resulting formalism ``Parameterised Post-Newtonian Cosmology'' (PPNC) \cite{Sanghai_2015, Sanghai_2017, Sanghai_2019}. This is a bottom-up approach to cosmology that explicitly allows for the presence of non-linear, inhomogeneously distributed matter, while carefully discarding the assumptions of asymptotically flat spacetime and negligible time-variation of the cosmological background that exist within the classical PPN formalism \cite{Will_1993}. Our approach generalises the concept of the ``slip'' and ``effective Newton's constant'' from previous attempts at creating a parameterized post-Friedmannian framework for testing gravity in cosmology \cite{Hu_2007, Hu_2008, Amin_2008, Skordis_2009, Baker_2011, Baker_2013}, and links these quantities, together with the background and time-dependendent parts of the linear-order gravitational field, to parameters familiar from the classical PPN approach. Here we develop this formalism by constructing a parameterised momentum constraint equation.

To begin our study of a theory-independent momentum constraint equation, let us consider a congruence of timelike curves with tangent vector $u^{\mu}$. We can now project the Ricci identities, $2\nabla_{[\mu}\nabla_{\nu]} u^{\rho} = R^{\rho}_{\:\mu\nu\sigma}u^{\sigma}$, and manipulate the result to obtain \cite{Ellis_1999}
\begin{equation}\label{1+3momentumconstraint}
D^{\nu}\sigma_{\mu\nu} - \frac{2}{3}D_{\mu}\Theta + g_{\mu\sigma}\eta^{\sigma\nu\rho}\left[D_{\nu}\omega_{\rho} + 2a_{\nu}\omega_{\rho}\right] = G_{\nu\rho}u^{\nu}h^{\rho}_{\: \mu},
\end{equation}
where $h_{\mu\nu} = g_{\mu\nu} + u_{\mu}u_{\nu}$ projects in the directions orthogonal to $u^{\mu}$, $\eta_{\sigma\nu\rho}$ is the volume form on the orthogonal hypersurfaces, $D$ is the spatial projection of $\nabla$, and $G_{\mu\nu} = R_{\mu\nu} - \frac{1}{2} g_{\mu\nu} R$ is the Einstein tensor. The kinematic quatities $\Theta$, $\sigma_{\mu\nu}$, $\omega_{\mu\nu}$ and $a^{\mu}$ are respectively the expansion, shear, vorticity and acceleration associated with $u^{\mu}$, which can be seen to be related to the momentum density in Eq. (\ref{1+3momentumconstraint}) through the term involving $G_{\mu\nu}$, once a set of field equations has been specified \cite{Ellis_1999}. This constraint must be satisfied in order for the theory to have a complete set of initial data, and is the relativistic generalization of the Newtonian requirement that the curl of the gradient of 3-velocities must vanish.

Equation (\ref{1+3momentumconstraint}) is the momentum constraint written in a theory-independent form, for any metric theory of gravity, and requires only a relationship between $G_{\mu \nu}$ and the stress-energy tensor $T_{\mu \nu}$ in order to be fully specified. In the context of Friedmann cosmology, we can choose coordinates such that $u^{\mu} = \left(u^0, 0,0,0\right)$, in which case Eq. (\ref{1+3momentumconstraint}) takes the more familiar form
\begin{equation}\label{cosmology1+3momentumconstraint}
-\frac{1}{a}\left[2\left(\mathcal{H}'-\mathcal{H}^2\right)\hat{B}_i + \frac{1}{2}\hat{\nabla}^2\hat{B}_i + 2\left(\hat{\Psi}'+\mathcal{H}\hat{\Phi}\right)_{,i}\right] = G_{0\rho}u^{0}h^{\rho}_{\: i} \, ,
\end{equation}
where we have chosen to write this equation in conformal time, and in longitudinal gauge. We have also assumed that the Robertson-Walker geometry is spatially flat. The scalars $\hat{\Phi}$ and $\hat{\Psi}$ are respectively the perturbations to the time-time and spatial parts of the metric, and $\hat{B}_i$ is the divergenceless vector perturbation to the time-space components (see below for proper definitions of these quantities). The quantity $\mathcal{H}=a'/a$ is the conformal Hubble rate. Our goal in this paper is to find versions of the scalar and divergenceless vector parts of this equation that can be written in terms of the PPNC parameters (to be explained in more detail below).

The plan of the paper is as follows: In Section \ref{sec2} we present a review of the required approximation schemes used in gravitational theory, as well as the PPN and PPNC constructions. The PPNC framework is then extended to include vector perturbations in Section \ref{sec:PNsmallscaleperturbationsFRW}, and the small-scale ($\lesssim 100 h^{-1}$ Mpc) limit of the parameterised momentum constraint equation is derived. In Section \ref{sec:largescales}, we consider very large scales in cosmology, and consider what this means for the parameterised momentum constraint. Our equations are then exemplified with a variety of test theories in Section \ref{sec:exampletheories}, before we conclude in Section \ref{sec:conclusions}.

We use Greek letters to denote spacetime indices, and Latin letters for spatial indices. We set $c = G = 1$ throughout. Commas represent partial derivatives, semicolons represent covariant derivatives, and dots and primes denote partial derivatives with respect to time and conformal time respectively. We will also use the convention that spatial indices on perturbed quantities are raised and lowered with a Kronecker delta, such that e.g. $B_{i,i} = B_{i,}^{\phantom{i,}i}= \delta^{ij} B_{i,j}$.

\section{Parameterised Post-Newtonian Cosmology}
\label{sec2}

In this section we will introduce the formalism in which our framework is constructed, which is a combination of post-Newtonian gravitational physics and cosmological perturbation theory. We will start by considering the essential features of both of these expansions, before moving on to describe the PPN formalism. We will then describe how the PPN formalism has been extended for use in cosmology, and recap some relevant results from previous papers.

\subsection{Weak-field expansions in cosmology}\label{sec:weakfields}

All of the expansions that we will use in this paper are ``weak field'', in the sense that there exist coordinate systems in which the metric can be written as
\begin{equation}
g_{\mu\nu} = g^{(0)}_{\mu\nu} + h_{\mu\nu}\, , \qquad {\rm where} \qquad g^{(0)}_{\mu \nu} \sim 1 \quad {\rm and} \quad h_{\mu\nu} \ll 1. 
\end{equation}
The components $g^{(0)}_{\mu\nu}$ will be referred to as the metric of the ``background'', which will be taken to correspond to either a Minkowski or Robertson-Walker geometry, and $h_{\mu \nu}$, which we will refer to as a ``perturbation''

The weak-field treatment is common to both cosmological perturbation theory and post-Newtonian expansions, and is justified by the leading-order part of the gravitational fields of all astrophysical objects except black holes and neutron stars being $U \ll 1$ (in geometrised units). On the other hand, cosmological perturbation theory and post-Newtonian expansions differ in the geometry of the assumed background, and in their treatment of the size of 3-velocities, $v$, of matter fields. In the former case the background is taken to be a Robertson-Walker geomety, and 3-velocities are taken to be of similar size in the perturbative expansion to gravitational potentials $U \sim v$. However, in the latter case the background is most commonly taken to be Minkowski space, and the leading-order part of the gravitational field is taken to be of the size $U \sim v^2$. These subtle differences have profound consequences, as we will now discuss.

\medskip
\noindent
{\it Cosmological Perturbation Theory}
\medskip

\noindent
On large scales in the universe, the simplest and most useful approach to model weak gravitational fields is to use cosmological perturbation theory \cite{Malik_2009, Kodama_1984, Bertschinger_2006}. In this approach the metric can be written
\begin{equation} \label{pertrw}
ds^2 = a(\h{\tau})^2 \left[ - (1-2 \h{\Phi}) d\h{\tau}^2 +\left( (1+2 \h{\Psi}) \delta_{ij} + \h{h}_{ij} \right) d\h{x}^i d\h{x}^j  + 2 \h{B}_i d\h{\tau} d\hat{x}^i \right] \, ,
\end{equation}
where $a(\h{\tau})$ is the scale factor, and where $\h{\Phi}$, $\h{\Psi}$, $\h{B}_i$ and $\h{h}_{ij}$ are all perturbations. This line-element is written in conformal time $\h{\tau}$, and we are free to choose a gauge such that $\h{B}_i$ is divergenceless and $\h{h}_{ij}$ is transverse and tracefree.

The perturbative order-of-smallness of all fields in this approach are taken to be similar, including the fluctuations in the density contrast, $\delta$, and the 3-velocities of matter fields, $v^i$, such that
\begin{equation}
\h{\Phi} \sim \h{\Psi} \sim \h{B}_i \sim \h{h}_{ij} \sim \delta \sim v^i  \ll 1 \,.
\end{equation}
The field equations of any theory of gravity can then be used to find the equations for the constraint and evolution equations for the background quantities, and subsequently those of all first and higher-order perturbations. This approach is highly flexible, and results in equations that are easy to solve and valid on a wide range of spatial and temporal scales. It does, however, have some drawbacks.

The principal among these is the fact the density constrast is required to be perturbatively small, and that the 3-velocity of matter fields is expected to remain as small as the amplitude of gravitational potentials. Neither of these things is true when we consider scales $\lesssim 100 \, {\rm Mpc}$ in the real Universe, where we can observe density contrasts $\delta\sim1$ or greater on scales $\lesssim 10 \, {\rm Mpc}$, and where we typically have $v^2 \sim \h{\Phi}$. This failure means that we cannot use cosmological perturbation theory to reliably model the gravitational interaction on scales $\lesssim 10 \, {\rm Mpc}$, and that we consequently face a challenge if we wish to try and use it to relate any parameterised framework for gravity in cosmology to results that we might obtain, for example, from experiments in the Solar System. This, together with the reliance on a set of field equations in which to perform the required perturbation theory, makes it very hard to conceive of a theory-independent parameterised framework for constraining gravity on all scales using cosmological perturbation theory alone.

\medskip
\noindent
{\it Post-Newtonian Theory}
\medskip

\noindent
In contrast to cosmological perturbation theory, post-Newtonian expansions do {\it not} assume that fluctuations in the mass density are small. This extra freedom is allowed as post-Newtonian theory assumes that gravity is not only weak-field, but also changing slowly with time, such that the time derivative of any quantity associated with matter or gravitational fields is small compared to the corresponding spatial derivatives of that quantity, i.e. such that
\begin{equation}
\frac{\frac{\partial}{\partial t}}{\frac{\partial}{\partial x}} \sim \vert v \vert  \ll 1.
\end{equation}
The slow-motion requirement is problematic for cosmology, as the Hubble flow increases in proportion to distance, and approaches $\sim 1$ on the scale of the horizon. It is probably for this reason that post-Newtonian expansions are usually specified using perturbations of Minkowski space:
\begin{equation}\label{pertm}
ds^2 = -(1-2 \Phi) dt^2 + (1+2 \Psi)\delta_{ij} dx^i dx^j + 2 B_{i} dt dx^i \, ,
\end{equation}
where in this expression we have $\Phi\sim \Psi \sim v^2$ and $\vert B_{i} \vert \sim v^3$, such that the vector gravitational potentials are smaller in magnitude than their scalar counterparts (the transverse and tracefree tensor perturbations are smaller still, so have been neglected).

While the line-element given in Eq. (\ref{pertm}) cannot be used to directly describe an entire cosmology, it can be safely applied within a region of space-time that is small compared to the cosmological horizon, so long as the Hubble flow velocity within that region is of order $v \ll 1$ (if this is not the case, then the slow motion requirement is violated).  By considering many such regions next to each other, one can then construct a viable cosmological model \cite{Sanghai_2015, Sanghai_2017}. This requires applying appropriate boundary conditions between each of the regions, which themselves allow the large-scale cosmological dynamics to emerge from the post-Newtonian-expanded gravitational fields. This is a construction known as ``post-Newtonian cosmology'', and has been investigated thoroughly in the context of Einstein's equations \cite{Sanghai_2015, Sanghai_2016}.

\medskip
\noindent
{\it Post-Newtonian Cosmology}
\medskip

\noindent
The link between the cosmological space-time that emerges in post-Newtonian cosmological modelling, and the perturbed Minkowski space in Eq.  (\ref{pertm}), can be made explicit by the following coordinate transformations:
\begin{eqnarray}\label{statictoexpandingtransformation1}
t &= \hat{t}+\frac{a^2 H}{2}\hat{r}^2 + T(\hat{t},\hat{\mathbf{x}}) + \mathcal{O}(v^5) \\\label{statictoexpandingtransformation2}
x^i &= a\,\hat{x}^i\left[1+\frac{a^2 {H}^2}{4}\hat{r}^2\right] + \mathcal{O}(v^4)\,,
\end{eqnarray}
where $T$ is an as-yet-unspecified gauge function of order $v^3$, ${H} \equiv \dot{a}/a$ is the Hubble parameter of the scale factor $a(\hat{t})$, and $\hat{r}^2 \equiv \delta_{ij} \h{x}^i \h{x}^j$. Under such a transformation, the line-element (\ref{pertm}) can be directly transformed into the form of the perturbed Robertson-Walker geometry (\ref{pertrw}), as long as we take
\begin{eqnarray}\label{expandingvsstatic}
{\Phi} &= \h{\Phi}+\frac{\ddot{a} \, a}{2}\hat{r}^2 \\\label{expandingvsstatic2}
\Psi &= \hat{\Psi} - \frac{\dot{a}^2}{4}\hat{r}^2 \\ \label{expandingvsstatic3}
B_i &= \hat{B}_i - 2 \dot{a} \, \hat{x}^j\delta_{ij}\left(\hat{\Phi}+\hat{\Psi}\right) - a \, \dot{a} \, \ddot{a} \, \hat{r}^2\hat{x}^j\delta_{ij} + \frac{1}{a}T_{,i}\, ,
\end{eqnarray}
and $\h{h}_{ij}=0$, and subsequently transform to conformal time. 

This demonstrates a direct isometry between the perturbed Minkowski space in which post-Newtonian gravity is usually formulated, and the perturbed Robertson-Walker geometries that are better suited to cosmology. As long as the coordinate patches of neighbouring regions overlap, which can be arranged by a suitable choice of $a(\hat{t})$, we can then consider this coordinate system to span the entire cosmology, and therefore to act as our `background'. This formulation of post-Newtonian gravity allows the gravitational fields of highly non-linear density contrasts to be consistently modelled, and simultaneously allows the Friedmann equations of the ``background'' to be extracted from them. It is therefore ideal for creating a unified framework for testing gravity in both isolated astrophysical systems, and in cosmology on the very largest scales.

\subsection{Parameterised post-Newtonian formalism}\label{sec:PPN}

The parameterised post-Newtonian (PPN) formalism is a method of constraining gravity using experimental and observational data {\it without} specifying a particular set of field equations or fundamental action for the underlying theory. This approach is based on the post-Newtonian expansion outlined above, and has proven itself to be extremely successful at providing a framework within which to understand gravity in a theory-independent fashion. Here we will spell out some of the crucial features of this approach, as relevant for our study.

The crucial first step in the classic PPN approach is to specify the perturbations to the metric (\ref{pertm}) in terms of matter fields and coupling parameters. This is typically done as follows \cite{Will_1993}:
\begin{eqnarray} \label{classicPPN0}
\nabla^2 \Phi &=& - 4 \pi \alpha \, \rho \,, \qquad {\rm} \qquad \nabla^2 \Psi = -4 \pi \gamma \, \rho 
\end{eqnarray}
and
\begin{eqnarray}\label{classicPPN}
\nabla^2 B_{i} &=& 8 \pi \Bigg[\alpha+\gamma+\frac{1}{4} \alpha_1 \Bigg] \rho \, v_i
-\Bigg[\alpha+\alpha_2-\zeta_1+2\xi\Bigg] \dot{U}_{,i} + \nabla^2{\varphi}^{\rm PF}_i 
\end{eqnarray}
where the Newtonian gravitational potential is defined implicitly by $\nabla^2 U \equiv - 4 \pi \, \rho$, and where the `preferred frame potential' is such that 
\begin{eqnarray}
\nabla^2 {\varphi}^{\rm PF}_i &=& 2 \pi {\alpha_1} {w}_i \, \rho +2 \alpha_2 {w}^j U_{,ij}\,,
\end{eqnarray}
where ${w}^j$ is the velocity of the PPN system with respect to the preferred frame of the theory, if one exists (to be explained in more detail later). 
There are six parameters appearing in these expressions: $\{\alpha, \gamma, \alpha_1, \alpha_2, \zeta_1 ,\xi \}$. The value of each of these should be understood to vary from theory to theory, but they can also be understood simply as coupling parameters for the gravitational potentials.

There are a number of comments that one could make about the parameterisation described above. Of most immediate relevance for our study is that the appearance of these parameters as coupling strengths for the graviational potentials is due to the structure of the post-Newtonian expansion itself, which means that the equations we will need to solve in any theory of gravity will take the form of Poisson-like equations. The simple form of the operators in such equations makes a specification in terms of a limited number of source terms possible, with the unknown parameters simply inserted as constants of proportionality. It is these equations that replace and remove the need for field equations of a particular gravitational theory, and that therefore allow a theory-independent interpretation of gravitational phenomena.

The appearance of $\{\alpha, \gamma, \alpha_1, \alpha_2, \zeta_1 ,\xi \}$ in the particular combinations in which they appear in Eqs. (\ref{classicPPN}) is so that they appear in global conservation laws in a simple way \cite{Will_1993}, and hence so that they can be associated with particular degrees of freedom in the space of theories of gravity (as outlined in Table 1). We note that in ``fully conservative'' theories of gravity, in which there are no violations of momentum or angular momentum conservation, and no preferred frame effects, we should have $\alpha_1= \alpha_2=\zeta_1 =0$. Finally, we note that Eqs. (\ref{classicPPN}) are written in ``post-Newtonian gauge'', which at the level of perturbations we are considering corresponds to a choice of gauge in which $h_{ij}$ is diagonal.

\subsection{Parameterised post-Newtonian cosmology}\label{sec:PNcosmology}

Let us now discuss how the PPN approach must be modified for application in cosmology, and what it looks like after the transformations (\ref{statictoexpandingtransformation1}) and (\ref{statictoexpandingtransformation2}).

To begin this, let us first note that the pressure of matter fields need not be neglected at leading order when considering the cosmological context, but that on the spatial scales on which post-Newtonian expansions can be applied (i.e. $\lesssim 100 \, {\rm Mpc}$) it must be effectively spatially constant \cite{Goldberg_2017b}. This is of particular importance for the inclusion of dark energy. Furthermore, we also note that the coupling parameters in Eq. \eqref{classicPPN} must be taken to be functions of cosmological time, in order to produce a consistent parameterized cosmological model \cite{Sanghai_2017}, i.e. 
\begin{equation}
\{\alpha, \gamma, \alpha_1, \alpha_2, \zeta_1 ,\xi \} \;\; \rightarrow \;\; \{\alpha (\hat{t}), \gamma (\hat{t}), \alpha_1 (\hat{t}), \alpha_2 (\hat{t}), \zeta_1 (\hat{t}),\xi(\hat{t}) \} \, .
\end{equation}
The reason for this is that gravitational couplings in alternative theories of gravity, in general, are allowed to be functions of additional background degrees of freedom, which themselves can change over cosmological time scales.

Taking these points properly into account, and performing the transformations (\ref{statictoexpandingtransformation1}) and (\ref{statictoexpandingtransformation2}) so that our space-time metric takes the form of a perturbed FRW metric, one finds that the appropriate Friedmann equations are given by the following set \cite{Sanghai_2016}:
\begin{eqnarray}
\mathcal{H}^2 &= \frac{8\pi\gamma}{3} \, \bar{\rho}a^2 - \frac{2\gamma_c a^2}{3} \label{parametrisedfriedmanneqns1} \\ 
\mathcal{H}' &= -\frac{4\pi\alpha}{3} \, \bar{\rho}a^2 + \frac{\alpha_c a^2}{3} \, ,\label{parametrisedfriedmanneqns2}
\end{eqnarray}
where $\mathcal{H}=a'/a$, $\bar{\rho}$ is the average energy density of matter in the Universe, and where primes denote differentiation with respect to conformal time, $\hat{\tau}$. The reader will note two extra parameters in these equations that do not appear in the classic PPN parameterisation: $\alpha_c$ and $\gamma_c$. These two terms must be added linearly to the right-hand side of the two equations in (\ref{classicPPN}) in order to be able to consistently include dark energy and the gravitational effects of the time variation of any extra degrees of freedom in a theory. They must satisfy the integrability condition
\begin{equation}\label{integrabilitycondition}
4 \pi \, \bar{\rho} \left(\alpha-\gamma +\frac{{\rm d}  \gamma}{{\rm d}  \ln a}  \right) = {\left(  \alpha_c+2 \gamma_c +\frac{{\rm d}  \gamma_c}{{\rm d}  \ln a}\right)} \, ,
\end{equation}
and must in general also be functions of cosmological time, such that $\alpha_c=\alpha_c(\hat{t})$ and $\gamma_c=\gamma_c(\hat{t})$. Eqs. (\ref{parametrisedfriedmanneqns1}) and (\ref{parametrisedfriedmanneqns2}) constitute the background contributions to the Hamiltonian constraint and Raychuadhuri equations, respectively. 

\begin{table}[t!]\label{PPNtable}
\begin{center}
\begin{tabular}{|p{5.5cm}||p{0.5cm}|p{4.5cm}|p{4.5cm}|}
\hline & & & \\[-10pt]
{\it Physical effect} & {\it } & {\it Observational constraint} & {\it Constraint on derivative} \\[5pt]
\hline\\[-15pt] \hline & & & \\[-10pt]
Effective Newton's constant  & $\alpha$ & $1$ & $0 \pm 0.01$ \cite{Uzan_2003} \\[5pt]
Spatial curvature parameter & $\gamma$ & $1 + \left(2.1 \pm 2.3\right)\times 10^{-5}$ \cite{Bertotti_2003} & $0 \pm 0.1$ \cite{Joudaki_2017} \\[5pt]
Preferred location parameter & $\xi$ & $0 \pm 3.9 \times 10^{-9}$ \cite{Shao_2013} & -- \\[5pt]
Conservation of momentum & $\zeta_1$ & $0 \pm 2 \times 10^{-2}$ \cite{Will_2014} & -- \\[5pt] 
Preferred frame parameters & $\alpha_1$ $\alpha_2$ & $\left(-0.7\pm 1.8 \right)\times 10^{-4}$ \cite{Shao_2012} $\left(1.8\pm 5.0\right) \times 10^{-5}$ \cite{Shao_2012}  & -- \newline -- 
\\[5pt]
Cosmological parameters & $\alpha_c$ $\gamma_c$ & $\left(2.07\pm 0.03\right)H_0^2$ \cite{Planck_2020} $\left(-1.04\pm 0.02\right)H_0^2$ \cite{Planck_2020} & $\left(0.12 \pm 0.25\right)H_0^2$ \cite{Planck_2020} $\left(-0.06\pm 0.12\right)H_0^2$ \cite{Planck_2020} \\[5pt]
\hline
\end{tabular}
\caption{The coupling parameters that appear at leading-order in the PPNC test metric, adapted for cosmology. An observational constraint on the derivative of a parameter $p$ refers to the constraint on $ \mathrm{d}p/\mathrm{d}\ln{a}$. 
}\label{tab1}
\end{center}
\end{table}

The leading-order perturbations to these equations have been investigated in Ref. \cite{Sanghai_2019}, where it was shown that they could be written as
\begin{eqnarray}
&\hspace{-25pt}
\frac{1}{3}\hat{\nabla}^2 \hat{\Psi} - \mathcal{H}^2\hat{\Phi} - \mathcal{H} \hat{\Psi}'   = -\frac{4\pi }{3} \, \mu \, \delta\! \rho \, a^2   \label{pert1} \\[3pt]
&\hspace{-25pt}
\frac{1}{3}\hat{\nabla}^2 \hat{\Phi} + 2{\mathcal{H}'}\hat{\Phi} +  \mathcal{H}\hat{\Phi}' + \hat{\Psi}'' +\mathcal{H} \hat{\Psi}'   = -\frac{4\pi }{3} \, \nu \, \delta\! \rho \, a^2 \, ,
 \label{pert2}
\end{eqnarray}
where $\delta \rho = \rho - \bar{\rho}$ is the perturbation to the energy density, and where $\mu$ and $\nu$ are in general functions of cosmological time and spatial scale. By comparing to the direct transformation of the classic PPN equations under Eqs. (\ref{statictoexpandingtransformation1}) and (\ref{statictoexpandingtransformation2}), we find that on scales $\lesssim 100 \, {\rm Mpc}$ we should expect \cite{Sanghai_2019}
\begin{eqnarray} \label{small}
\lim_{k \rightarrow \infty} \mu &= \gamma \qquad {\rm and} \qquad \lim_{k \rightarrow \infty} \nu = \alpha \, .
\end{eqnarray}
These equations are required for the theory to be correctly described by the classic PPN approach on small spatial scales. On the other hand, by employing the separate universe approach \cite{Bertschinger_2006}, we find that the super-horizon limit of adiabatic perturbations should be described by \cite{Sanghai_2019}
\begin{eqnarray} \label{large1}
\lim_{k \rightarrow 0} \mu &= \gamma - \frac{1}{3} \frac{{\rm d} \gamma}{{\rm d}  \ln a}+ \frac{1}{12 \pi \bar{\rho}} \frac{{\rm d} \gamma_c}{{\rm d} \ln a} \\ \label{large2}
\lim_{k \rightarrow 0} \nu &= \alpha - \frac{1}{3} \frac{{\rm d}  \alpha}{{\rm d}  \ln a}+ \frac{1}{12 \pi \bar{\rho}} \frac{{\rm d}  \alpha_c}{{\rm d}  \ln a} \, .
\end{eqnarray}
This behaviour is required for consistency with the effective Friedmann equations (\ref{parametrisedfriedmanneqns1})-(\ref{parametrisedfriedmanneqns2}). The limiting behaviour given in Eqs. (\ref{small})-(\ref{large2}) shows that the parameters $\mu$ and $\nu$ must approach a predictable scale-invariant form on both small and large spatial scales, and that in general we should expect the values of these to be different in these two limits. In all cases, however, the limiting behaviour of $\mu$ and $\nu$ must be a function of the set of extended-PPN parameters $\{\alpha, \gamma, \alpha_c , \gamma_c\}$.

It is the purpose of the present paper to extend the set of equations outlined above, to include not only the Hamiltonian constraint and the Raychaudhuri equation, but also the momentum constraint equation. The momentum constraint in cosmology is usually throught of as being comprised of two parts; a scalar equation that adds a constraint to the set described above, and a divergenceless vector equation that contains the ``frame-dragging'' potential $\hat{B}_i$. We will investigate both parts of the momentum constraint equation, on both small spatial scales (where a post-Newtonian expansion can be performed), and on very large scales (where the parameterised Friedmann equations can be applied). 

\section{The Momentum Constraint Equation on Small Scales}\label{sec:PNsmallscaleperturbationsFRW}

In this section we will use the transformations from Eqs. (\ref{statictoexpandingtransformation1}) and (\ref{statictoexpandingtransformation2}) to derive the momentum constraint equation for parameterised post-Newtonian perturbations around a Friedmann background, as they occur in the metric (\ref{pertrw}). As the post-Newtonian expansion on which these equations are based is expected to be valid on scales $\lesssim 100 \, {\rm Mpc}$, this will give us the ``small scale'' limit of the general parameterised momentum constraint equation. We will start by considering this equation in conservative theories of gravity. These are theories in which global energy, momentum and angular momentum are conserved to post-Newtonian order in asymptotically flat spacetime \cite{Will_1993}. In terms of PPN parameters, they correspond to the values $\alpha_1=\alpha_2=\zeta_1=0$. We will then generalise to non-conservative theories.

\subsection{Conservative theories}\label{subsec:conservativetheories}

Let us first use the relationships between perturbations to Minkowski and Robertson-Walker geometries, specified by Eqs. \eqref{expandingvsstatic}-\eqref{expandingvsstatic3}, to derive the constraint equation satisfied by the cosmological vector perturbation $\h{B}_i$. Assuming we want our cosmological perturbations to be in longitudinal gauge \cite{Clifton_2020}, we must have that $\h{B}_{i, i}=0$, where the derivative is taken with respect to the $\hat{x}^i$ coordinates. Using Eq. \eqref{expandingvsstatic3}, we find that this condition is satisfied if and only if the gauge function in Eq. (\ref{statictoexpandingtransformation1}) satisfies
\begin{equation}
\hat{\nabla}^2 T = a^2 B_{i,i} + 6a \dot{a} \left(\hat{\Phi}+\hat{\Psi}\right) + 2a \dot{a} \left(\hat{\Phi}+\hat{\Psi}\right)_{,i} \hat{x}^i+ 5a^2 \dot{a} \ddot{a}\, \hat{r}^2 \, . 
\end{equation}
In this equation, and henceforth, we will use the convention that spatial derivatives on a quantity are with respect to the set of coordinates with which that quantity is defined (i.e. so derivatives of hatted quantities are taken with respect to $\h{x}^i$, and unhatted quantities are differentiated with respect to $x^i$). 

We now wish to operate on Eq. \eqref{expandingvsstatic3} with the Laplacian $\hat{\nabla}^2$, and substitute in the expression above for $T$, in order to find the following expression for the left-hand side of the momentum constraint on small scales:
\begin{eqnarray} \label{smom1}
\hspace{-1cm}& 
\frac{1}{2a } \h{\nabla}^2 \h{B}_i + 2 \left(\dot{\h{{\Psi}}} + {H} \h{\Phi} \right)_{,i}
\\\hspace{-1cm}= & 
\frac{1}{2} a \left( \nabla^2 B_i - B_{j,ji} \right) 
+ {H} \hat{x}_i \hat{\nabla}^2 \left( \hat{\Phi} + \hat{\Psi} \right) -{H} \hat{x}^{j}\left( \hat{\Phi}+ \hat{\Psi} \right)_{,ij} +2 \dot{\h{{\Psi}}}_{,i}-2 {H}  \hat{\Psi}_{, i} \nonumber \, .
\end{eqnarray}
To go further we now need an expression for $B_i$, which we take from solving Eq. \eqref{classicPPN} with  $\alpha_1=\alpha_2=\zeta_1=0$ to obtain
\begin{equation} \label{PPNti}
B_i = -2 (\alpha + \gamma) V_i + \left( \frac{1}{2} \alpha + \xi \right) \dot{\chi}_{,i} + B^{\rm extra}_{i} \, ,
\end{equation}
where $\nabla^2 V^{i} \equiv -4 \pi \rho v^{i}$ and $\nabla^2 \chi \equiv -2 U$ implicitly define the vector potential $V^i$ and the super-potential $\chi$, and where we have added the extra term $B^{\rm extra}_{i}$ to account for any extra contributions that may need to be added to the right-hand side of Eq. \eqref{classicPPN} in order to make it suitable for use in cosmology.

Taking the required derivatives of $B_i$, and substituting back into Eq. \eqref{smom1}, gives us
\begin{eqnarray} \label{smom2}
\hspace{-30pt}
&
\frac{1}{2a} \h{\nabla}^2 \h{B}_i + 2 \left( \dot{\h{{\Psi}}}+{H} \h{\Phi} \right)_{,i}
\\\hspace{-30pt} =& 4 \pi a (\alpha+ \gamma) \rho v_i + a (\alpha+\gamma) V_{j,ji} +  {H} \hat{x}_i \hat{\nabla}^2 (\hat{\Phi} + \hat{\Psi}) \nonumber \\ \nonumber \hspace{-30pt} &-  {H} \hat{x}^{j} ( \hat{\Phi} + \hat{\Psi} )_{, ij}
+2 \dot{\h{{\Psi}}}_{,i} - 2 {H} \hat{\Psi}_{, i} + \frac{1}{2}a \left( \nabla^2 B^{\rm extra}_{i} - B^{\rm extra}_{j ,ji} \right) \, . 
\end{eqnarray}
To proceed further it is useful to re-write the term containing the factor $V_{j,ji}$. We do this by splitting the velocity $v_i$ into a background part (due to the Hubble flow) and a peculiar velocity, such that $v^i = H {x}^i + \delta {v}^i$. This then allows us to split $V^i$ into components due to the background and peculiar velocities, such that $V_i = \bar{V}_i + \delta {V}_i$, where
\begin{equation}\label{Vhat}
\nabla^2 \bar{V}_i \equiv -4\pi\rho H {x}_i \qquad {\rm and} \qquad \nabla^2 \delta {V}_i \equiv -4\pi\rho\, \delta{v}_i \, .
\end{equation}
The former of these implicit definitions allows for the solution $\bar{V}^{i} = H {x}^{i} U + H \chi_{,i}$, where the derivative should be understood to be with respect to the $x^i$ coordinates.

Making a similar split of the Newtonian potential $U$ into contributions from the cosmological background and perturbation, i.e. taking $\rho= \bar{\rho} + \delta \rho$, allows us to write
\begin{equation}\label{Ubar_and_deltaU}
\nabla^2 U = -4 \pi \rho = - 4 \pi \bar{\rho} - 4 \pi \delta \rho \equiv \nabla^2 \bar{U} + \nabla^2 \delta U \, ,
\end{equation}
where the last equality provides implicit definitions for $\bar{U}$ and $\delta U$. We note that these allow us to write down the solution for the background part of the Newtonian potential as $\bar{U} = -{2 \pi} \bar{\rho} r^2/3$, and the cosmological perturbations to the Robertson-Walker geometry as $\hat{\Phi} = \alpha \delta U$ and $\hat{\Psi} = \gamma \delta U$. These expressions, together with the continuity equation $\dot{\rho} + (\rho v^i)_{,i} = 0$, allow us to derive the useful identities 
\begin{eqnarray} \nonumber \hspace{-1.5cm}
\alpha \, \frac{\partial \;}{\partial x^i } \, \delta {V}_{i}&= -\dot{\hat{\Phi}}-{H}\hat{\Phi}+\frac{\dot{\alpha}}{\alpha}\hat{\Phi}
\qquad {\rm and} \qquad
\gamma \, \frac{\partial \;}{\partial x^i }\, \delta {V}_{i} = -\dot{\hat{\Psi}}-{H}\hat{\Psi}+\frac{\dot{\gamma}}{\gamma}\hat{\Psi} \, ,
\end{eqnarray}
which can be compared with the identity $V_{i,i} = - \dot{U}$ that frequently occurs in the classic approach to post-Newtonian gravity \cite{Will_1993}. 

Using all of these results in Eq. (\ref{smom2}), it follows that we can write the momentum constraint in the cosmological geometry as
\begin{eqnarray}\label{vectornablainPF}
\hspace{-75pt}&
\frac{1}{2a} \h{\nabla}^2 \h{B}_i + 2 \left(\dot{\h{{\Psi}}}+ {H} \h{\Phi} \right)_{,i}
 \\
\hspace{-75pt}= &4\pi\left(\alpha+\gamma \right)\rho \,\delta{v}_i \, a - \Big[\Big(\dot{\hat{\Phi}}-\dot{\hat{\Psi}}\Big) - {H}\left(\hat{\Phi}-\hat{\Psi}\right) - \frac{\dot{\alpha}}{\alpha}\hat{\Phi}-\frac{\dot{\gamma}}{\gamma}\hat{\Psi}\Big]_{,i}  
+ \frac{1}{2} a\left[\nabla^2 B^{\mathrm{extra}}_{i} - B^{\mathrm{extra}}_{j,ji}\right] \, . \nonumber \label{vectornablainPF1.5}
\end{eqnarray}

This can be straightforwardly split into a scalar part
\begin{eqnarray}\label{vectornablainPF2}
\hspace{-60pt}&  2  {\h{{\Psi}}}_{,i}'+2 \mathcal{H} \h{\Phi}_{,i}  =4\pi\left(\alpha+\gamma \right) \left[ \rho \hat{v}_i \right]^{\rm S} \, a^2 -\Big[\Big({\hat{\Phi}}-{\hat{\Psi}}\Big)' - \mathcal{H}\left(\hat{\Phi}-\hat{\Psi}\right) - \frac{{\alpha}'}{\alpha}\hat{\Phi}-\frac{{\gamma}'}{\gamma}\hat{\Psi}\Big]_{,i}   ,
\end{eqnarray}
and a divergenceless vector part
\begin{eqnarray}\label{vectornablainPF3}
\hspace{-60pt}&\hat{\nabla}^2 \hat{B}_i = 8\pi\left(\alpha+\gamma \right) \left[\rho \hat{v}_i\right]^{\rm V} \, a^2 + \h{\nabla}^2 \hat{B}_i^{\rm extra} \, ,
\end{eqnarray}
where $\h{\nabla}^2 \hat{B}_i^{\rm extra}\equiv a^2(\nabla^2 B^{\mathrm{extra}}_{i} - B^{\mathrm{extra}}_{j,ji})$ is manifestly a divergence-free vector, and where $\h{v}^i=d\h{x}^i/d\h{\tau} = \delta v^i $. In each of these two expressions we have used the superscipts S and V on parentheses to indicate that the we intend this to correspond to the scalar or divergenceless-vector part of the object within. We have also converted all time derivatives in these expressions into conformal time, so that they appear in the most familiar form for use in cosmology. The reader may note that although the PPN parameter $\xi$ is allowed to be non-zero in this class of theories, it appears in neither the scalar nor the vector part of our parameterised momentum constraint equation.

This is our first step towards the momentum constraint on small cosmological scales. Let us now consider how these expressions are generalised in non-conservative theories of gravity, which can exhibit violation of global conservation laws at post-Newtonian order, and which can display preferred-frame effects. 

\subsection{Non-conservative theories}\label{subsec:preferredframe}

In non-conservative theories of gravity the PPN parameters $\{\alpha_1, \alpha_2, \zeta_1 \}$ are allowed to be non-zero. This can be accounted for by using Eq. \eqref{classicPPN} without setting any of the PPN parameters to zero. Following a similar process to the one used above, this results in
\begin{eqnarray}\label{vectornablainPF4}
\hspace{-40pt}&\hat{\nabla}^2 \hat{B}_i = 8\pi\left(\alpha+\gamma+\frac{\alpha_1}{4} \right) \left[\rho \hat{v}_i\right]^{\rm V} \, a^2 
+ 2 \pi \alpha_1 a^2 \mathcal{H} \left[\rho\, \hat{x}^i\right]^{\rm V}
+ \h{\nabla}^2 \hat{B}_i^{\rm extra} \, ,
\end{eqnarray}
where we have again allowed for extra terms to be added by including $\h{\nabla}^2 \hat{B}_i^{\rm extra}$, and where we have assumed that we are in the preferred frame (if one exists), such that $\varphi_i^{\rm PF}=0$. The reader may note that only $\alpha_1$ is retained in the cosmological version of this equation, with $\alpha_2$ and $\zeta_1$ both being removed by the imposition that $\h{B}_i$ is divergenceless. 

The reader may also note that there is a highly anomolous term on the right-hand side of this equations: $ 2 \pi \alpha_1 a^2 \mathcal{H} \rho\, \hat{x}^i$. This term appears problematic as it depends linearly on the distance from the origin of coordinates. As we have not specified a particular configuration of matter, or any symmetries beyond those of the background, it is hard to see how such a term could possibly be permitted, even in theories with preferred frames. We will therefore remove it by altering Eq. (\ref{classicPPN}) so that it takes the new form
\begin{equation*} 
\hspace{-40pt}\nabla^2 B_{i} = 8 \pi (\alpha+\gamma ) \rho \, v_i
+2 \pi \alpha_1  \rho \, \delta v_i
-(\alpha+\alpha_2-\zeta_1+2\xi) \dot{U}_{,i} +\nabla^2 B_i^{\rm extra} \, ,\nonumber
\end{equation*}
where $\delta{v}_i=v_i-H x^i$, and where we are again temporarily taking $\varphi_i^{\rm PF} = 0$. This adjustment corresponds to allowing only the peculiar component of the 3-velocity $v^i$ to source the part of the equation that couples with $\alpha_1$, and results in
\begin{eqnarray}\label{vectornablainPF5}
\hspace{-40pt}&\hat{\nabla}^2 \hat{B}_i = 8\pi\left(\alpha+\gamma+\frac{\alpha_1}{4} \right) \left[\rho \hat{v}_i\right]^{\rm V} \, a^2 
+ \h{\nabla}^2 \hat{B}_i^{\rm extra} \, .
\end{eqnarray}
It is conceivable that similar changes may need to be made for the term that couples with $\alpha_2$, as this is also a preferred-frame parameter, but as this parameter does not appear in the cosmological equations at the order we are studying we will not concern ourselves with it here.

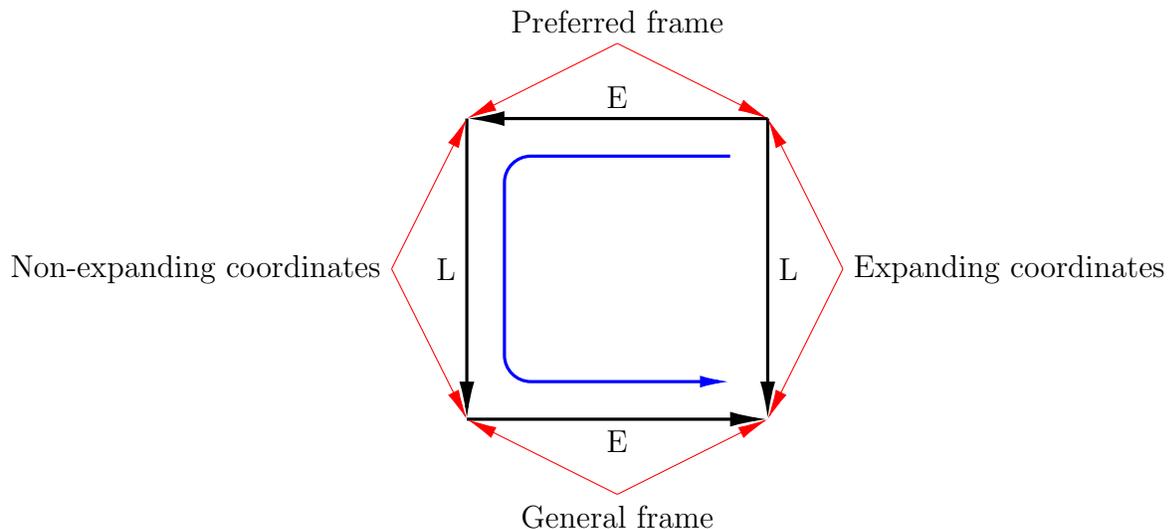
\begin{figure}[t]
\begin{tikzpicture}
\coordinate (A) at (0,0);
\coordinate (B) at (0,4);
\coordinate (C) at (4,4);
\coordinate (D) at (4,0);
\coordinate (E) at (0,2);
\coordinate (F) at (2,4);
\coordinate (G) at (4,2);
\coordinate (H) at (2,0);
\coordinate (I) at (-1,2);
\coordinate (J) at (2,5);
\coordinate (K) at (5,2);
\coordinate (L) at (2,-1);
\draw[very thick,{Latex[length=5mm, width=2mm]}-](A)--(B);
\draw[very thick,{Latex[length=5mm, width=2mm]}-](B)--(C);
\draw[very thick,{Latex[length=5mm, width=2mm]}-](D)--(C);
\draw[very thick,{Latex[length=5mm, width=2mm]}-](D)--(A);
\draw[-{Latex[length=4mm, width=1.6mm]},color=red](I)--(A);
\draw[-{Latex[length=4mm, width=1.6mm]},color=red](I)--(B);
\draw[-{Latex[length=4mm, width=1.6mm]},color=red](J)--(B);
\draw[-{Latex[length=4mm, width=1.6mm]},color=red](J)--(C);
\draw[-{Latex[length=4mm, width=1.6mm]},color=red](K)--(C);
\draw[-{Latex[length=4mm, width=1.6mm]},color=red](K)--(D);
\draw[-{Latex[length=4mm, width=1.6mm]},color=red](L)--(D);
\draw[-{Latex[length=4mm, width=1.6mm]},color=red](L)--(A);
\draw[-{Latex[length=4mm, width=1.6mm]}, very thick, color=blue, rounded corners=10pt] (3.5,3.5) -- (0.5,3.5) -- (0.5,0.5) -- (3.5,0.5);
\node[left] at (E) {L};
\node[above] at (F) {E};
\node[right] at (G) {L};
\node[below] at (H) {E};
\node[left] at (I) {Non-expanding coordinates};
\node[above] at (J) {Preferred frame};
\node[right] at (K) {Expanding coordinates};
\node[below] at (L) {General frame};

\end{tikzpicture}

\caption{A schematic of the transformations between perturbed Minkowski and perturbed Robertson-Walker geometries, and tranformations between the preferred frame and general frames. Lorentz transformations are labelled L, and E denotes transformations between non-expanding and expanding backgrounds.}
\label{fig:coordsystemsPF}
\end{figure}

We will now turn our attention to the form this equation takes if we transform away from the preferred frame, so that our coordinate system is in motion with respect to it. We will generate the new form of this equation by performing a Lorentz boost in the perturbed Minkowski description of the space-time. The corresponding perturbed Robertson-Walker descriptions, before and after the boost, can then be determined by using the transformations from Eqs. \eqref{statictoexpandingtransformation1} and \eqref{statictoexpandingtransformation2}. This process is displayed schematically in Fig. \ref{fig:coordsystemsPF}, where the transformation from perturbed Robertson-Walker geometry in the preferred frame to the general frame is indicated by the black arrow from the top-right corner to the bottom-right corner, and which is equivalent to the three transformations around the other sides of the square, collectively denoted by the blue arrow. 

We have already discussed the transformations between expanding and static backgrounds, corresponding to the top and bottom of the square in Fig. \ref{fig:coordsystemsPF}, in previous sections. The Lorentz boost between two different coordinate systems covering Minkowski space, corresponding to the left side of the square, are given by the standard expressions:
\begin{eqnarray}
t &\rightarrow {t} \left( 1+ \frac{1}{2} w^2 + \frac{3}{8} w^4 \right) + \left( 1+ \frac{1}{2} w^2 \right) {x}^i w_i +O(v^5) \times {t} \\
x^i &\rightarrow {x}^i +\left( 1+ \frac{1}{2} w^2 \right)  {t}\,  w^i+\frac{1}{2} {x}^j w_j w^i +O(v^4) \times {x}^i \, .
\end{eqnarray}
The effects of these on the perturbations to Minkowski space are
\begin{eqnarray}
\Phi \rightarrow \Phi \, , \qquad \Psi \rightarrow \Psi \, ,  \qquad {\rm and} \qquad
B_i \rightarrow B_i - 2 w_i (\Phi+\Psi) \, .
\end{eqnarray}
Transforming back to expanding coordinates after performing this boost, and ensuring the gauge function $T$ is chosen to maintain longitudinal gauge, we find that 
\begin{equation}\label{smallscalevectorNC}
\hspace{-20pt}\hat{\nabla}^2 \hat{B}_i = 8\pi\left(\alpha+\gamma\right)\left[\rho \hat{v}_i\right]^{\rm V}a^2 + 2\pi\alpha_1\left[\rho\left(\hat{v}_i+\hat{w}_i\right)\right]^{\rm V}a^2 + \h{\nabla}^2 \hat{B}_i^{\rm extra} \,  ,
\end{equation}
where $\hat{w}^i$ is the coordinate velocity relative to the preferred frame in the perturbed Robertson-Walker description of the space-time. This is the general form of the vector part of the parameterised momentum constraint equation, written in a form suitable for cosmology. The reader may note that $\hat{w}^i$ does not enter into the terms that couple with $\alpha$ and $\gamma$, which is a direct consequence of these terms having no preferred frame, as per their usual interpretation in the classic PPN formalism. They may also note that in order to avoid the presence of another problematic term, which would be linear in $\hat{x}^i$, we have had to adjust the preferred frame potential $\varphi_i^{\rm PF}$ so that it too depends only on the velocity relative to the Hubble flow, $\delta {w}_i = w_i - H x^i$.

\section{The Momentum Constraint on the Largest Scales}\label{sec:largescales}

We now focus on extending our parameterised equations up to super-horizon scales. Much of this will rely on a separate universe approach, as pioneered in a theory-independent way by Bertschinger \cite{Bertschinger_2006}. 
This approach has allowed the Raychaudhuri and Hamiltonian constraint equations for super-horizon perturbations to be obtained within the PPNC formalism \cite{Sanghai_2019}, and we now consider how it can be applied to the momentum constraint equation on those scales.

\subsection{Boosts, and the scalar momentum constraint equation}\label{subsec:largescalescalar}

Let us start by considering the effect of a boost in the coordinates of the expanding Robertson-Walker space, such that we transform to new coordinates
\begin{eqnarray}
\hat{\tau}^* &= \gamma(\hat{v})\left[\hat{\tau} - \hat{v} \hat{x}\right] = \hat{\tau} - \hat{v}\hat{x} + \frac{\hat{v}^2}{2}\hat{\tau} + ... \\
\hat{x}^* &= \gamma(\hat{v})\left[\hat{x} - \hat{v} \hat{\tau}\right] = \hat{x} - \hat{v}\hat{\tau} + \frac{\hat{v}^2}{2}\hat{x} + ... 
\, ,
\end{eqnarray}
and $\hat{y}^* = \hat{y}$ and $\hat{z}^* = \hat{z}$, where $\hat{v}$ is the velocity of the boost in the $\hat{x}$-direction. The conformal part of the metric is unchanged by this transformation, while the scale factor becomes
\begin{equation}
a^2(\hat{\tau}^*) \simeq a^2(\hat{\tau})\left[1 - 2\hat{v}\mathcal{H}\hat{x}\right].
\end{equation}
Hence, the line-element for our geometry becomes
\begin{equation}
ds^2 = a^2(\hat{\tau})\left[-\left(1-2\mathcal{H}\theta\right)\mathrm{d}\hat{\tau}^2+ \left(1-2\mathcal{H}\theta\right)\left\lbrace\mathrm{d}\hat{x}^2 + \mathrm{d}\hat{y}^2 + \mathrm{d}\hat{z}^2\right\rbrace \right],
\end{equation}
where $\theta = \hat{v}\hat{x}$ is the scalar velocity potential. This is equivalent to the following pair of scalar perturbations:
\begin{equation}\label{scalarsvstheta}
\hat{\Phi} = \mathcal{H}\theta \qquad {\rm and} \qquad \hat{\Psi} = -\mathcal{H}\theta \, . 
\end{equation}
Constructing the left-hand side of the momentum constraint equation therefore gives that on super-horizon scales we must have
\begin{equation}\label{superhorizonscalar}
2 \hat{\Psi}' + 2 \mathcal{H}\hat{\Phi} = 2 \left(\mathcal{H}^2 - \mathcal{H}'\right) \theta \, ,
\end{equation}
which can equivalently be expressed as
\begin{equation}\label{superhorizonscalar2}
2 \hat{\Psi}'_{,i} + 2 \mathcal{H}\hat{\Phi}_{,i} = \frac{8 \pi}{3} (\alpha + 2 \gamma) \bar{\rho} \hat{v}^{\rm S}_i \,a^2- \frac{2}{3} (\alpha_c+ 2 \gamma_c) \hat{v}^{\rm S}_i\, a^2 \, ,
\end{equation}
where we have used the parameterised Friedmann equations (\ref{parametrisedfriedmanneqns1}) and (\ref{parametrisedfriedmanneqns2}), and generalised this expression to an arbitrary direction by taking $\theta_{,i} = \hat{v}^{\rm S}_i$, where $\hat{v}^{\rm S}_i$ is the scalar part of the matter peculiar velocity field $\hat{v}_i$. This result can be seen to be consistent with the scalar Hamiltonian constraint and Raychaudhuri equations, derived using Bertschinger's separate universe treatment, and the large-scale momentum conservation equation, which we take as evidence of the validity of our approach.

We find that Eq. \eqref{superhorizonscalar2} and the results from Section \ref{sec:PNsmallscaleperturbationsFRW} can be written together in a single equation as
\begin{equation}\label{generalparametrisedscalareqn}
\boxed{\hat{\Psi}'_{,i} + \mathcal{H}\hat{\Phi}_{,i} = 4\pi \mu \left[\rho \hat{v}_i\right]^{\rm S} a^2 + \mathcal{G} \mathcal{H} \hat{\Psi}_{,i} \phantom{\Big]}} \,,
\end{equation}
where $\mathcal{G}=\mathcal{G}\left(\hat{\tau},k\right)$ is assumed to be a smooth function with limits
\begin{eqnarray}
\label{G1largek} \lim_{k \rightarrow \infty}\mathcal{G} &=& \frac{{\rm d}\ln \gamma}{{\rm d}\ln a} +\frac{\alpha-\gamma}{\gamma} 
\qquad {\rm and} \qquad
\label{G1smallk} \lim_{k \rightarrow 0}\mathcal{G} = 0 \,,
\end{eqnarray}
and where $\mu$ is given by Eqs. (\ref{large1}) and (\ref{large2}). In order to derive this result, we have made use of Eq. \eqref{vectornablainPF2} and the post-Newtonian-expanded matter continuity equation (valid on small scales), as well as Eqs. \eqref{parametrisedfriedmanneqns1}, \eqref{integrabilitycondition} and \eqref{superhorizonscalar} for the super-horizon limit. It is intended that in the super-horizon limit the combination $\rho \hat{v}_i$ should be understood as approaching $\bar{\rho} \hat{v}_i$, as in this limit the density contrast is assumed to be perturbatively small. We note that the function $\mathcal{G}$ vanishes on all scales for the case of GR with a cosmological constant, in which case $\alpha = \gamma=1$. This will not be the case in general though, and for modified theories of gravity it is expected that $\mathcal{G} \neq 0$ on small scales.

\subsection{Rotations, and the vector momentum constraint equation}\label{subsec:largescalevector}

\begin{figure}[htb!]
    \centering
    \includegraphics[width=1\textwidth]{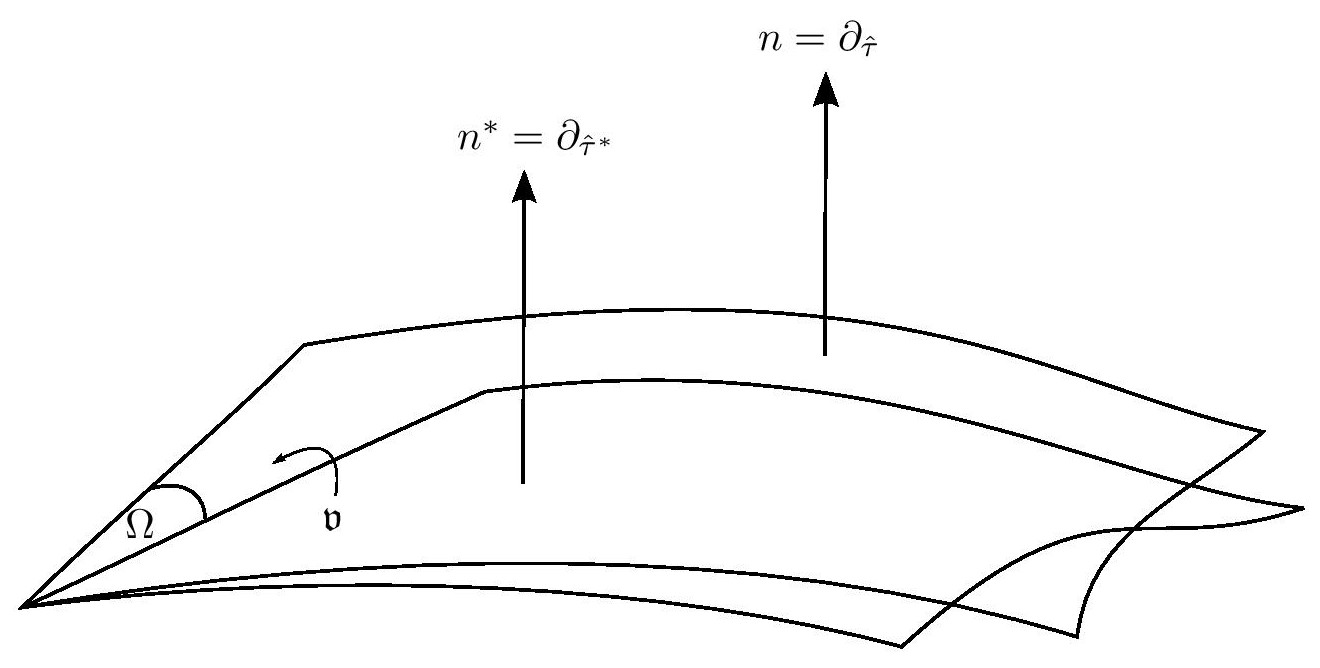}
    \caption{Rotating coordinates by a time-dependent angle $\Omega$ induces the divergenceless vector perturbation $\mathfrak{v}_i$ in the fluid. Depicted here are space-like hypersurfaces of constant $\hat{\tau}$.}
    \label{fig:rotation}
\end{figure}

In order to construct a divergenceless vector version of the momentum constraint equation for super-horizon scales, let us now consider the case where we rotate our spatial coordinates, rather than boosting them. This will produce an apparent vortical motion in the fluid that fills the space-time, as illustrated in Figure \ref{fig:rotation}. 

In order to induce this perturbation we rotate coordinates by the angle $\Omega=\Omega(\hat{\tau})$, such that
\begin{eqnarray}
\hat{x}^* &= \h{x} \cos{\Omega} +\h{y} \sin{\Omega} \\ 
\hat{y}^* &= \h{y} \cos{\Omega}-\h{x} \sin{\Omega} \, ,
\end{eqnarray}
with $\h{z}^*=\h{z}$ and $\hat{\tau}^*=\hat{\tau}$. Putting this into the line-element of a spatially flat Robertson-Walker geometry gives us
\begin{eqnarray}
\mathrm{d}s^2 =a^2(\hat{\tau})\left[-\mathrm{d}\hat{\tau}^2+\mathrm{d}\hat{x}^2+\mathrm{d}\hat{y}^2+\mathrm{d}\hat{z}^2\right]  +2a^2(\hat{\tau})\frac{\mathrm{d}\Omega}{\mathrm{d}\h{\tau}}\left(\hat{y}\mathrm{d}\hat{x}-\hat{x}\mathrm{d}\hat{y}\right)\mathrm{d}\hat{\tau}\, ,
\end{eqnarray}
where we have taken $\Omega = \Omega (\hat{\tau})$ and expanded to leading order in $d\Omega/d\hat{\tau}$. Comparing this to a linearly-perturbed Robertson-Walker geometry allows us to identify that the rotating coordinates are in longitudinal gauge, and that we have induced a divergenceless-vector perturbation
\begin{equation}\label{generalDVhorizonscale}
\hat{B}_i = \frac{{\rm d}\Omega}{{\rm d}\h{\tau}} \left(\hat{y}, -\hat{x},0\right) = -\hat{\mathfrak{v}}_i\, ,
\end{equation}
where $\hat{\mathfrak{v}}_i$ is the divergenceless vector part of the fluid's 3-velocity in the rotating coordinates. Clearly there is nothing special about the direction of the axis of rotation in this example, so we expect the result $\hat{B}_i = - \hat{\mathfrak{v}}_i$ to be valid in general.

Combining our result with the relevant prefactor allows us to write down a straightforward equation for the divergenceless vector part of the super-horizon momentum constraint equation:
\begin{equation}\label{horizonfullconstraint}
2 \left(\mathcal{H}'-\mathcal{H}^2\right)\hat{B}_i 
=-2 \left(\mathcal{H}'-\mathcal{H}^2\right) \hat{\mathfrak{v}}_i \, , 
\end{equation}
or, equivalently, using Eqs. (\ref{parametrisedfriedmanneqns1}) and (\ref{parametrisedfriedmanneqns2}), as 
\begin{equation} \label{veceq}
2 \left(\mathcal{H}'-\mathcal{H}^2\right)\hat{B}_i = \frac{8 \pi}{3} (\alpha + 2 \gamma) \bar{\rho}\, \hat{\mathfrak{v}}_i \,a^2- \frac{2}{3} (\alpha_c+ 2 \gamma_c) \hat{\mathfrak{v}}_i \, a^2 \, .
\end{equation}
This is of exactly the same form as the scalar part of the super-horizon momentum constraint equation (\ref{superhorizonscalar2}), which gives us confidence in its validity.

We can now see that it is possible to write Eq. (\ref{veceq}) together with the divergenceless vector equation from Section \ref{sec:PNsmallscaleperturbationsFRW} in the unified form
\begin{equation}\label{generalparametrisedvectoreqn}
\hspace{-2cm} \boxed{ 2 \left(\mathcal{H}'-\mathcal{H}^2\right)\hat{B}_i+\frac{1}{2} \hat{\nabla}^2\hat{B}_i  = 8\pi (\mu+\mathcal{Q}) \left[\rho \hat{v}_i\right]^{\rm V} a^2  +  
 \alpha_1 \, \pi \left[\rho \hat{w}_i\right]^{\rm V} a^2
}\,  ,
\end{equation}
where $\mu$ is again given by Eqs. (\ref{large1}) and (\ref{large2}), and the coupling function $\mathcal{Q}$ has the limits
\begin{eqnarray}
\label{G2largek} \lim_{k \rightarrow \infty}\mathcal{Q} &=& \frac{\alpha - \gamma}{2} + \frac{\alpha_1}{8} 
\qquad {\rm and} \qquad
\label{G2smallk} \lim_{k \rightarrow 0}\mathcal{Q} = 0\,.
\end{eqnarray}
In deriving these equations we have made use of Eqs. \eqref{smallscalevectorNC} and \eqref{horizonfullconstraint}, as well the background equations \eqref{parametrisedfriedmanneqns1} and \eqref{parametrisedfriedmanneqns2}, and the integrability condition \eqref{integrabilitycondition}. Again, the quantities $\rho \hat{v}_i$ should be understood to reduce to $\bar{\rho} \hat{v}_i$ on large scales, when the density contrast becomes perturbatively small. 
The new function $\mathcal{Q}$ vanishes identically in GR, when $\alpha=\gamma=1$ and $\alpha_1=0$, but is not expected to be zero on small scales in generic modified theories of gravity. The reader may also note that we have set $\hat{B}_i^{\rm extra}=0$, as we find it not to be required for the theories of gravity below.

Equations \eqref{generalparametrisedscalareqn} and \eqref{generalparametrisedvectoreqn} are the central results of this paper. In what follows, we will demonstrate their applicability using some specific example theories.

\section{Example Theories}\label{sec:exampletheories}

Having derived our parameterised momentum constraint equations, we will now show how they work for some example theories of modified gravity and dark energy. This requires determining the PPNC parameters for each theory, and then demonstrating that inputting these into our equations results in the correct small and large-scale limits of their weak-field theory.

Before we proceed with the worked examples, it will prove useful to collect together some general results. For this, we note that for adiabatic perturbations the gauge-invariant entropy perturbation $S_{XY}$ between two scalars $X$ and $Y$ vanishes:
\begin{equation}\label{adiabaticcondition}
S_{XY} := \mathcal{H}\left(\frac{\delta X}{\bar{X}'} - \frac{\delta Y}{\bar{Y}'}\right) = 0.
\end{equation}
We also note that the conservation equation $\nabla_{\nu}T^{0\nu} = 0$ implies that on super-horizon scales we have
\begin{equation}\label{largescaledeltarho1}
\bar{\rho}' + 3\mathcal{H}\bar{\rho} = 0 \qquad {\rm and} \qquad 
\delta\rho' + 3\mathcal{H}\delta\rho + 3\bar{\rho}\hat{\Psi}' = 0 \,,
\end{equation}
which imply that
\begin{equation}\label{largescaledeltarho3}
\frac{\delta\rho}{\bar{\rho}'} = \frac{\hat{\Psi}}{\mathcal{H}} \, , \qquad {\rm and \;\; therefore} \qquad
\frac{\delta X}{\bar{X}'} = \frac{\hat{\Psi}}{\mathcal{H}},
\end{equation}
for scalars $X$ that appear in the theory.
It will also be useful to note the scalar part of $\nabla_{\nu}T^{i\nu} = 0$ on large scales:
\begin{equation}\label{Eulerequation}
\theta' + \mathcal{H}\theta - \hat{\Phi} = 0 \,, \qquad {\rm which \;\; implies} \qquad \hat{\Psi}=- \mathcal{H} \theta \, .
\end{equation}
Let us now consider our example theories, in increasing order of mathematical complexity.

\subsection{Quintessence}\label{sec:quintessence}

Let us start with quintessence, which is a scalar field, $\phi$, that minimally couples to gravity. The full action of these theories is
\begin{eqnarray}
I &= \frac{1}{16\pi} \int \mathrm{d}^4 x \sqrt{-g}\left[R -\frac{1}{2}g^{\alpha\beta}\nabla_{\alpha}\phi\nabla_{\beta}\phi - V(\phi)\right] + I_m\left(\psi,g_{\mu\nu}\right) \,,
\end{eqnarray}
where $\psi$ denotes matter fields. The field equations are
\begin{equation}\label{quintessenceEinsteinEquations}
G_{\mu\nu} = 8\pi T_{\mu\nu} + 8 \pi \left( g_{\mu\nu}\left[-\frac{1}{2}g^{\alpha\beta}\nabla_{\alpha}\phi\nabla_{\beta}\phi - V(\phi)\right] + \nabla_{\mu}\phi \nabla_{\nu}\phi  \right) \,,
\end{equation}
and $\square \phi = dV/d\phi$. Performing a post-Newtonian expansion about Minkowski space gives the following PPNC parameters:
\begin{equation}
\hspace{-1cm}
\alpha = \gamma = 1, \quad
\gamma_c = -4\pi\left(\frac{\left(\bar{\phi}'\right)^2}{2a^2} + V(\bar{\phi})\right), \: \: \alpha_c = -8\pi\left(\frac{\left(\bar{\phi}'\right)^2}{a^2} - V(\bar{\phi})\right),
\end{equation}
where $\bar{\phi}$ is the time-dependent background value of $\phi$, and $\alpha_1 = \alpha_2= \xi = \zeta_1=0$.
Let us now investigate the form of the perturbation equations in these theories.

\vspace{0.5cm}
\noindent
{\it Small scales}:
Applying these parameter values above to Eqs. \eqref{generalparametrisedscalareqn} and \eqref{generalparametrisedvectoreqn}, we find
\begin{equation}\label{quintessencesmallscale0i}
\frac{1}{2}\hat{\nabla}^2\hat{B}_i + 2\left[\hat{\Psi}+\mathcal{H}\hat{\Phi}\right]_{,i} = 8\pi\rho\hat{v}_i a^2 \, .
\end{equation}
It now remains to show that this is the same equation one would obtain from performing a direct post-Newtonian expansion of Eq. \eqref{quintessenceEinsteinEquations}. For this, we can note that $\square \phi = dV/d\phi$ can be expanded to give
\begin{equation}
\frac{1}{a^2}\left(\bar{\phi}'' + 3 \mathcal{H} \bar{\phi}'\right) = - \frac{dV (\bar{\phi})}{d\phi}  \qquad {\rm and} \qquad \hat{\nabla}^2 \delta \phi = 0 \, ,
\end{equation}
where we have separated out the leading-order part of this equation into its background and inhomogeneous parts, using $\phi= \bar{\phi} + \delta \phi$ and taking $\delta \phi \sim v^2$. As the inhomogeneous equation has no source terms, this implies that the leading-order part of the quintessence field in the post-Newtonian expansion must be homogeneous, which in turn means that all contributions to leading-order part of the $0i$-field equation from the scalar field must vanish on small scales \footnote{See Ref. \cite{Goldberg_2017b} for a more detailed discussion of this phenomenon.}. We are therefore led to an equation that is identical to \eqref{quintessencesmallscale0i}, from our direct analysis of the field equations \eqref{quintessenceEinsteinEquations}, which verifies our parameterised equation for this example.

\vspace{0.5cm}
\noindent
{\it Large scales}:
For super-horizon scales, the parameterised scalar equation \eqref{generalparametrisedscalareqn} becomes
\begin{equation}
\hat{\Psi}'+\mathcal{H}\hat{\Phi} = 4\pi\bar{\rho}a^2\theta + 4\pi\bar{\phi}^{\prime 2}\theta \, ,
\end{equation}
where $\hat{v}_i=\theta_{,i}$. This can be compared to the equation for scalar super-horizon perturbations, derived directly from the field equations \eqref{quintessenceEinsteinEquations}:
\begin{equation}
\hat{\Psi}'+\mathcal{H}\hat{\Phi} = 4\pi\bar{\rho}a^2\theta - 4\pi\bar{\phi}'\delta\phi.
\end{equation}
It can be seen that these two equations are identical provided that $\delta\phi = -\theta\bar{\phi}'$, which can be obtained from the adiabatic condition (\ref{largescaledeltarho3}) and the Euler equation (\ref{Eulerequation}). The parameterised scalar momentum constraint on super-horizon scales is therefore identical to what is obtained from directly expanding the field equations \eqref{quintessenceEinsteinEquations} .

The divergenceless vector part is even simpler: the super-horizon limit of our general result \eqref{generalparametrisedvectoreqn} follows immediately from the field equations. We therefore have that our parameterised momentum constraint equations correctly reproduces all of the results that one would obtain from directly dealing with the quintessence model of dark energy, in both the scalar and divergenceless vector sectors of the theory, and on both large and small limits. We therefore have our first explicit verification of its validity. 
\subsection{Brans-Dicke theory}\label{sec:Brans-Dicke}

Let us now consider Brans-Dicke theory, which is a scalar-tensor theory specified by the action
\begin{equation}
I = \frac{1}{16\pi}\int \mathrm{d}^4 x \sqrt{-g}\left[\varphi R - \frac{\omega}{\varphi}\nabla^{\mu}\varphi\nabla_{\mu}\varphi\right] + I_m\left(\psi, g_{\mu\nu}\right).
\end{equation}
The field equations of this theory are
\begin{equation}\label{BDTmetric}
\hspace{-1cm}
G_{\mu\nu} + \frac{g_{\mu\nu}}{\varphi}\left[\Box \varphi + \frac{\omega}{2\varphi}\nabla^{\alpha}\varphi\nabla_{\alpha}\varphi\right] - \frac{1}{\varphi}\nabla_{\mu}\nabla_{\nu}\varphi - \frac{\omega}{\varphi^2}\nabla_{\mu}\varphi\nabla_{\nu}\varphi = \frac{8\pi}{\varphi}T_{\mu\nu} \, ,
\end{equation}
and
\begin{equation}\label{BDTscalarfield}
\hspace{-1cm}
\Box\varphi = \frac{8\pi}{3+2\omega}T \, .
\end{equation}
The PPN parameters of this theory are 
\begin{eqnarray}\label{BransDickePPNparameters}
\alpha = \frac{4+2\omega}{3+2\omega} \, \frac{1}{\bar{\varphi}} \,, \quad
\gamma = \frac{2+2\omega}{3+2\omega} \, \frac{1}{\bar{\varphi}} \, ,  \quad {\rm and} \quad
\alpha_1 = \alpha_2 = \xi = \zeta_1 = 0 \, ,
\end{eqnarray}
with cosmological parameters
\begin{eqnarray}
\hspace{-2cm}
\alpha_c =\frac{1}{a^2} \left[ -\frac{\bar{\varphi}^{''}}{\bar{\varphi}} + \mathcal{H} \frac{\bar{\varphi}'}{\bar{\varphi}} - \omega\left(\frac{\bar{\varphi}'}{\bar{\varphi}}\right)^2 \right]\, , \quad
\gamma_c  = -\frac{1}{2 a^2}\left[\frac{\bar{\varphi}^{''}}{\bar{\varphi}}-\mathcal{H}\frac{\bar{\varphi}'}{\bar{\varphi}} + \frac{\omega}{2}\left(\frac{\bar{\varphi}'}{\bar{\varphi}}\right)^2\right] \,.
\end{eqnarray}
Here we have expanded the scalar field as $\varphi\left(\hat{\tau},\hat{\mathbf{x}}\right) = \bar{\varphi}\left(\hat{\tau}\right) + \delta\varphi\left(\hat{\tau},\hat{\mathbf{x}}\right)$. The FRW equations can be obtained by applying these PPNC parameter values to Eqs. \eqref{parametrisedfriedmanneqns1} and \eqref{parametrisedfriedmanneqns2}. Let us now consider the momentum constraint for this theory.

\vspace{0.5cm}
\noindent
{\it Small scales}: We can immediately write down the scalar part of the momentum constraint \eqref{generalparametrisedscalareqn} on small scales as
\begin{equation}\label{RHSBDTscalars}
2\left[\hat{\Psi}+\mathcal{H}\hat{\Phi}\right]_{,i} = \frac{8\pi}{\bar{\varphi}}\left[\rho \hat{v}_i\right]^{\rm S}a^2 - \frac{\hat{\Psi}'_{,i}}{1+\omega} + \frac{\mathcal{H}\hat{\Psi}_{,i}}{1+\omega} - \frac{\bar{\varphi}'}{\bar{\varphi}}\hat{\Phi}_{,i} - \frac{\bar{\varphi}'}{\bar{\varphi}}\hat{\Psi_{,i}},
\end{equation}
and the divergenceless vector part  \eqref{generalparametrisedvectoreqn} as
\begin{equation} \label{bdvec}
\hat{\nabla}^2\hat{B}_i = \frac{16\pi a^2}{\bar{\varphi}}\left[\rho \hat{v}_i\right]^{\rm V}\,.
\end{equation}
Let us now show that a direct post-Newtonian expansion of the $0i$-field equation (\ref{BDTmetric}) generates the same results.

Focusing on the scalar part of equation (\ref{BDTmetric}) gives
\begin{equation}\label{BDTmomentumconstraintscalar}
2\left[\hat{\Psi}'+\mathcal{H}\hat{\Phi}\right]_{,i} = \frac{8\pi}{\bar{\varphi}}\left[\rho \hat{v}_i\right]^{\rm S}a^2 + \frac{1}{\bar{\varphi}}\left[\mathcal{H}\delta\varphi-\hat{\Phi}\bar{\varphi}'-\delta\varphi'\right]_{,i} - \frac{\omega \bar{\varphi}'}{\bar{\varphi}^2}\delta\varphi_{,i}.
\end{equation}
To deal with the terms involving $\delta \varphi$, let us note that a post-Newtonian expansion of the scalar field equation \eqref{BDTscalarfield} tells us that
\begin{equation}
\hat{\nabla}^2 \delta \varphi = -\frac{8\pi}{3+2\omega}\, \delta\rho\, a^2 \, , \qquad {\rm which \;\; implies} \qquad \delta \varphi = \frac{\bar{\varphi}}{1+\omega} \,\hat{\Psi}\, .
\end{equation}
Using this result, it can be seen that Eq. \eqref{BDTmomentumconstraintscalar} readily reduces to Eq. \eqref{RHSBDTscalars}, which verifies our parameterised equation \eqref{generalparametrisedscalareqn} in this case. We have also verified that the divergenceless vector part of Eq. \eqref{BDTmetric} correctly reproduces (\ref{bdvec}), which follows straightforwardly as there are no direct contributions from the scalar field to the divergenceless vector part of the $0i$-field equation: it enters only through the combination $\alpha + \gamma=2/\bar{\varphi}$.

\vspace{0.5cm}
\noindent
{\it Large scales}: On super-horizon scales, the parameterised scalar equation \eqref{generalparametrisedscalareqn} can be written
\begin{equation}\label{superhorizonscalarBDTPPNC}
2\left[\hat{\Psi}'+\mathcal{H}\hat{\Phi}\right] = \frac{8\pi\bar{\rho}a^2}{\bar{\varphi}} \, \theta +  \left[-2\mathcal{H} \frac{\bar{\varphi}'}{\bar{\varphi}} + \frac{\bar{\varphi}^{''}}{\bar{\varphi}} + \omega\left(\frac{\bar{\varphi}'}{\bar{\varphi}}\right)^2\right]  \theta\, ,
\end{equation}
where we have made use of the background equation for the scalar field \eqref{BDTscalarfield}, which reads
\begin{equation}
\bar{\varphi}^{''} + 2\mathcal{H}\bar{\varphi} = \frac{8\pi\bar{\rho}a^2}{3+2\omega} \, .
\end{equation}
For adiabatic perturbations, Eqs. \eqref{largescaledeltarho3} and \eqref{Eulerequation} give
\begin{equation}\label{adiabaticscalarfield}
\frac{\delta\varphi}{\bar{\varphi}'} = -\theta \, , \qquad {\rm and} \qquad
\frac{\bar{\varphi}^{''}}{\bar{\varphi}} \, \theta= -\frac{\delta\varphi'}{\bar{\varphi}} \, .
\end{equation}
Substituting these results into Eq. \eqref{superhorizonscalarBDTPPNC}, we get
\begin{equation}\label{BDTsuperhorizonscalars}
2\left[\hat{\Psi}'+\mathcal{H}\hat{\Phi}\right] = \frac{8\pi\bar{\rho}a^2}{\bar{\varphi}}\, \theta + \frac{1}{\bar{\varphi}}\left[\mathcal{H}\delta\varphi - \hat{\Phi}\bar{\varphi}' - \delta\varphi'\right] - \frac{\omega\bar{\varphi}'}{\bar{\varphi}^2}\delta\varphi \,,
\end{equation}
which is precisely what is obtained by directly expanding the scalar part of the $0i$-field equation \eqref{BDTmetric}. The divergenceless vector part again agrees immediately with Eq. \eqref{generalparametrisedvectoreqn}, which verifies that our parameterised equations reproduce the results of Brans-Dicke theory exactly on both small and large scales, and in both the scalar and divergenceless-vector sectors of the theory.

\subsection{Vector-tensor theory}\label{sec:Vector-tensor}

Let us now show our parameterised momentum constraint also works in theories that contain a time-like vector field, $A^{\mu}$, as well as the metric. The gravitational action for such theories may be written as \cite{Will_1993}
\begin{equation}\label{VTaction}
I = \int \mathrm{d}^4 x \frac{\sqrt{-g}}{16\pi}\left[\left(1+\omega A_{\mu}A^{\mu}\right)R  -2 \omega A^{\mu}A^{\nu}R_{\mu\nu} 
+ \tau A_{\mu;\nu}A^{\mu;\nu}\right],
\end{equation}
where $\omega$ and $\tau$ are the coupling constants, and we have chosen to consider a simplified subclass of vector-tensor theories that nevertheless displays all the gravitational effects of interest in this paper. 
The field equations for these theories are
\begin{equation} \label{VTmetricEOM}
G_{\mu\nu} + \tau\Theta^{(\tau)}_{\mu\nu} + \omega\Theta^{(\omega)}_{\mu\nu} - 2\omega\Theta^{(\eta)}_{\mu\nu} = 8\pi T_{\mu\nu},
\end{equation}
where
\begin{eqnarray} \label{VT_Theta_tau}
\hspace{-2cm}
\Theta^{(\tau)}_{\mu\nu} &=& A_{\mu;\sigma}A_{\nu}^{\:\: ;\sigma} + A_{\sigma;\mu}A^{\sigma}_{\:\: ;\nu} - \frac{g_{\mu\nu}}{2}A_{\sigma;\rho}A^{\sigma;\rho} 
+ \left[A^{\sigma}A_{(\mu;\nu)} - A^{\sigma}_{\:\: ;(\mu}A_{\nu)} - A_{(\mu}A_{\nu)}^{\:\: ;\sigma}\right]_{;\sigma}\\\hspace{-2cm}
\label{VT_Theta_omega} \Theta^{(\omega)}_{\mu\nu} &=& RA_{\mu}A_{\nu} + A_{\rho}A^{\rho}G_{\mu\nu} - \left(A_{\rho}A^{\rho}\right)_{;\mu\nu} + g_{\mu\nu}\Box\left(A_{\rho}A^{\rho}\right) \\\hspace{-2cm}
\label{VT_Theta_eta} \Theta^{(\eta)}_{\mu\nu} &=& 2A^{\rho}A_{(\mu}R_{\nu)\rho} - \frac{g_{\mu\nu}}{2}A^{\rho}A^{\sigma}R_{\rho\sigma} - \left[A^{\rho}A_{(\mu}\right]_{;\nu)\rho} 
+ \frac{1}{2}\Box\left(A_{\mu}A_{\nu}\right) + \frac{g_{\mu\nu}}{2}\left(A^{\rho}A^{\sigma}\right)_{;\rho\sigma},
\end{eqnarray}
with the corresponding field equation for the vector being given by
\begin{equation}\label{VTvectorEOM}
2\omega A^{\nu}G_{\mu\nu} + \tau\Box A_{\mu} = 0.
\end{equation}
The PPN parameters for this theory are
\begin{eqnarray}\label{hellingsPPNCparamsalpha}
\hspace{-2.6cm}\alpha &=& \frac{2\left[\tau +\omega\bar{A}^2\left(8\omega-\tau\right)\right]}{\tau\left[2+\bar{A}^2\left(\tau-4\omega\right)-\omega\bar{A}^4\left(\tau-10\omega\right)\right]}\,,
\;\label{hellingsPPNCparamsgamma} \gamma= \frac{2\left(1-\omega\bar{A}^2\right)}{2+\bar{A}^2\left(\tau-4\omega\right)-\omega\bar{A}^4\left(\tau-10\omega\right)}\,,\\ 
\label{hellingsPFparameter}
\hspace{-2.6cm}\alpha_1 &=& \frac{16\tau}{2\tau+\bar{A}^2\left(2\tau\left(\tau+\omega\right)-\left(\tau+2\omega\right)^2\right)}-\frac{16\tau-2\omega\bar{A}^2\left(\tau-4\omega\right)}{2\tau+\tau\bar{A}^2\left(\tau-4\omega\right)-\omega\tau\bar{A}^4\left(\tau-10\omega\right)},
\end{eqnarray} 
and $\zeta_1 = \xi = 0$, and the cosmological parameters are
\begin{eqnarray}\label{hellingsalphac}
\hspace{-2.6cm}\nonumber \alpha_c &=& \frac{1}{a^2} \Bigg[ \left(\alpha-\frac{2}{2-\bar{A}^2\left(\tau-2\omega\right)}\right)\frac{\tau\left(2-\bar{A}^2\left(\tau-2\omega\right)\right)\left(\bar{A}^{''}+2\mathcal{H}\bar{A}'\right)}{\bar{A}^2\left(\tau-2\omega\right)} 
\end{eqnarray}
\begin{eqnarray}
\hspace{2cm}- \left(\alpha+\frac{6}{2-\bar{A}^2\left(\tau-2\omega\right)}\right)\frac{\tau\bar{A}^{'\,2}}{4}+ \frac{6\mathcal{H}\bar{A}\bar{A}'\left(\tau-2\omega\right)}{2-\left(\tau-2\omega\right)\bar{A}^2} \Bigg]\,,
\end{eqnarray}
\begin{eqnarray} \label{hellingsgammac}
\hspace{-2.6cm}\gamma_c &=& \frac{1}{a^2} \Bigg[ \left(\gamma - \frac{2}{2-\bar{A}^2 \left(\tau-2\omega\right)}\right)\frac{\tau\left(2-\bar{A}^2\left(\tau-2\omega\right)\right)\left(\bar{A}^{''}+2\mathcal{H}\bar{A}'\right)}{4\bar{A}^2\left(\tau-2\omega\right)} - \frac{\gamma \tau\bar{A}^{'\,2}}{4} \Bigg] \, ,
\end{eqnarray}
where $\bar{A}$ is the background value of $A_0$, and where time derivatives are with respect to the FRW conformal coordinate $\hat{\tau}$. 

\newpage
\noindent
{\it Small scales}: 
Let us focus first on the post-Newtonian regime. In this case we can write $$A_{\mu} = \left(\bar{A} + \delta A_0^{(2)}, \delta A_i^{(1)} + \delta A_i^{(3)}\right)\,,$$ where $\bar{A}\sim O(1)$, and superscripts indicate the perturbative order in $v$. It can immediately be noted that the vector field equation of motion gives $\hat{\nabla}^2\delta A^{(1)}_i = 0$, which with suitable boundary conditions implies $\delta A^{(1)}_i=\delta A^{(1)}_i(\hat{\tau})$. As $\delta A^{(1)}_i$ is spatially constant, it must necessarily be the derivative of a scalar, i.e. $\delta A^{(1)}_i = \delta A^{(1)\, {\rm S}}_i$.

We can now recognise that these theories have a preferred frame, which is picked out by the direction of the timelike vector field $A_{\mu}$.
To complete the full set of ingredients required to compute the momentum constraint on small scales in the PPNC framework, we need to add to the PPNC parameters the velocity $\hat{w}_i$ of expansion-comoving observers with respect to the preferred frame, and the cosmological divergenceless vector $\hat{B}^{\rm extra}_i$. We will now find those in turn. 

To determine the preferred-frame 3-velocity $\hat{w}_i$ that couples to $\alpha_1$, consider a local Lorentz boost from the preferred frame, in which $\hat{w}_i$ vanishes, to a generic frame, in which it does not. The ``preferred frame'' refers in the case of these vector-tensor theory to the frame picked out by a preferred time direction that is aligned with the timelike vector field, i.e. a frame constructed using the coordinates $\left(\tau^*, \mathbf{x}^*\right)$ in which $\delta A^{(1)}_i = 0$.

We can now perform the Lorentz transformation to the generic frame $\left(\tau,\mathbf{x}\right)$, which for ease of calculation we present as the inverse transformation:
\begin{eqnarray}
\tau^* &=& \gamma_w\left(\tau+\hat{w}_j x^j\right); \\
x^{*\,j} &=& \gamma_w \left(x^j+\hat{w}^j\tau\right).
\end{eqnarray}
Computing the transformation of the vector field components in the usual way, we find that $\delta A^{(1)}_i = \hat{w}_i \bar{A}$, and so the preferred-frame velocity $\hat{w}_i$ is directly related to the local perturbation to the vector field by 
\begin{equation}
\hat{w}_i = \frac{\delta A^{(1)}_i}{\bar{A}},
\end{equation}
which we recall from Section \ref{subsec:preferredframe} has no divergenceless vector part. 

To determine the ``extra'' cosmological contribution to the local vector perturbation, as in Eq. \eqref{smallscalevectorNC}, we expand the vector part of the $0i$-field equation about Minkowski spacetime, allowing for the time evolution of $\bar{A}$ (which is negligible in the classic PPN formalism), and look for the relevant additional term. This gives
\begin{equation}\label{B_i_extra_VT}
\hat{\nabla}^2\hat{B}^{\rm extra}_i = -\frac{2\tau}{1+\left(\tau+\omega\right)\bar{A}^2}\left[\bar{A}^{''}+2\mathcal{H}\bar{A}'\right]\delta A^{(1)\,{\rm V}}_i = 0,
\end{equation}
because $\delta A^{(1)}_i$ has no divergenceless vector part.
With all ingredients obtained, we can substitute back into the momentum constraint Eqs. \eqref{vectornablainPF2} and \eqref{smallscalevectorNC}, which reconstitutes the full momentum constraint for the small-scale metric perturbations in these theories.

\newpage
\noindent
{\it Large scales}:
On super-horizon scales, we know from our earlier analysis that the parameterised momentum constraint must be given by
\begin{equation}
2\left(\hat{\Psi}'+\mathcal{H}\hat{\Phi}\right)_{,i} - 2\left(\mathcal{H}^2-\mathcal{H}'\right)\hat{B}_i = 2\left(\mathcal{H}^2-\mathcal{H}'\right)\hat{v}_i.
\end{equation}
Using the parameter values from Eqs. \eqref{hellingsPPNCparamsalpha}--\eqref{hellingsgammac}, and simplifying them using the equation of motion for the vector field, we have that the coefficient of $\hat{v}_i$ on the right hand side is
\begin{equation}
2\left(\mathcal{H}^2-\mathcal{H}'\right) = \frac{8\pi\bar{\rho}a^2 + \tau \bar{A}'^{\, 2} - 2\mathcal{H}\bar{A}\bar{A}'\left(\tau-2\omega\right)}{1-\frac{\bar{A^2}}{2}\left(\tau-2\omega\right)}.
\end{equation}
This allows us to explicitly reconstruct the momentum constraint on large scales, which the reader may note is a vast simplification compared to a direct perturbation theory expansion of the field equations of these theories.

\section{Conclusions}\label{sec:conclusions}

We have extended the PPNC formalism, introduced in Refs. \cite{Sanghai_2017,Sanghai_2019}, by deriving a parameterised momentum constraint equation. This study has included gravitational potentials that result from theories with preferred frame effects, and has resulted in a parameterised scalar equation \eqref{generalparametrisedscalareqn} and a parameterised divergenceless-vector equation \eqref{generalparametrisedvectoreqn}.

Our parameterised equations are valid on scales where the density contrast is highly non-linear, as well as on super-horizon scales where terms in the field equations with time derivatives dominate. They require the introduction of only one new parameter: $\alpha_1(\hat{\tau})$, which is expected to be non-zero in non-conservative theories of gravity only, and which at the present time coincides with the PPN parameter of the same name. We find that the additional PPN parameters $\alpha_2$, $\zeta_1$ and $\xi$, which are present in the PPN equation for vector gravitational potentials, are not necessary for the cosmological version of the momentum constraint equation.

Ultimately, we expect the principal utility of parameterised frameworks in cosmology to come from their comparison to observational data. In this regard, the parameter $\alpha$ is likely to best constrained by the leading-order behaviour of matter, which in cosmology could include observations such as the growth rate of structure, and the matter power spectrum. This parameter is required to be precisely equal to one at the present time, by definition of units such that $G=1$ at this time, but could have a time dependence that will likely be best constrained using cosmological observations.

On the other hand, the parameter $\gamma$ is likely to best constrained with observations that depend on the paths of rays of light. In cosmology, this would include weak lensing and CMB lensing. Beyond leading-order, the parameter $\alpha_1$ could potentially also be constrained by such observations, as on scales where non-linear structures are present the vector potential $\hat{B}_i$ will appear in the next-to-leading order equations, at only one order higher in $v/c$ \cite{Thomas_2015,Milillo_2015}. Both of these parameters are well constrained at the present time, but once again it will likely be cosmological observations that will provide the tightest constraints on their time dependence.

Beyond cosmological observations, there is also the possibility of constraining the time dependence of our parameters using observations of gravitational waves emitted from compact binary inspirals. These events can be detected from systems at high cosmological redshifts, which means they present (in principle) an environment in which the time-variation of gravitational coupling constants could be constrained or detected. Indeed, it is expected that gravitational wave signals should be sensitive to the preferred-frame PPN parameters $\alpha_1$ and $\alpha_2$ \cite{Sampson_2013}. It would be an exciting prospect to use these newly detected systems to complement cosmological probes.

In future work we intend to complete the full system of equations required to describe parametrized equations in cosmology. This will require developing an understanding of the interpolation of the coupling functions $\mu$, $\nu$, $\mathcal{G}$ and $\mathcal{Q}$ between small and large scales \cite{Clifton_2021}, as well as complementing the existing equations with a ``slip equation'' (typically derived from the shear evolution equation, in most theories). We hope it will also include developing equations from the transverse-tracefree tensor sector of the parameterised frameworks, which are problematic to include in standard post-Newtonian expansions as they only couple to matter at relatively high orders, as well as higher-order scalar potentials.

Finally, we also intend to extend our formalism to incorporate theories that exhibit non-perturbative screening mechanisms \cite{Vainshtein_1972,Khoury_2004,Khoury_2010,Avilez_Lopez_2015} and Yukawa potentials \cite{Khoury_2004, Clifton_2008}. Such theories have more complicated post-Newtonian limits that cannot be written purely in terms of the standard set of PPN parameters \cite{Clifton_2011}, and so to include them in our system would require extending the original PPN formalism. This seems like a worthwhile endeavour, however, as these theories have provoked a lot of interest in the cosmology literature (indeed, they are often the {\it only} theories considered in many studies).

\section*{Acknowledgments}\label{sec:acknowledgments}

Several calculations in this paper were performed using the tensor algebra package xAct and its subpackages xPert \cite{Brizuela_2009}, xPand \cite{Pitrou_2013} and xPPN \cite{Hohmann_2021}. We thank Daniel Thomas and Pedro Carrilho for helpful discussions, and acknowledge support from the STFC under grant ST/P000592/1.

\appendix

\section{Separate universe approach for spatially flat backgrounds}\label{sec:separateuniverse}

The approach used by Bertschinger \cite{Bertschinger_2006} assumes that the initial Robertson-Walker geometry has non-zero curvature, and that this can be written as a second Robertson-Walker geometry with a spatial curvature that is equal to the first up to a factor that is perturbatively close to unity. Here we present a similar derivation, but instead under the assumption that the second geometry is spatially flat, as relevant for our presentation.

For this, consider a Robertson-Walker space-time:
\begin{equation}
\mathrm{d}s^2 = a^2(\hat{\tau})\left[-\mathrm{d}\hat{\tau}^2 + \frac{{\bf \mathrm{d}\hat{x}}^2}{1+\frac{K}{4} {\bf \hat{x}}^2}\right].
\end{equation}
Now let us assume the spatial curvature is small such that $K=\delta K \ll 1$, and perturb the coordinates such that $\tau \longrightarrow \hat{\tau} = \tau + A(\tau)$ and $x^i \longrightarrow \hat{x}^i = x^i\left(1+\beta(\tau)\right)$, where $A(\tau) \ll1$ and $\beta(\tau) \ll 1$. We then obtain
\begin{eqnarray}
\label{metricA}
\hspace{-1.9cm}\mathrm{d}s^2 &=& a^2(\tau) \Bigg[-\left(1+2A'+2\mathcal{H}A + 2 \frac{\delta K}{a} \frac{\partial a}{\partial K}\right)\mathrm{d}\tau^2 \\
\nonumber \hspace{-1.9cm}&&\qquad + 2\beta' x_i \mathrm{d}\tau\mathrm{d}x^i + \left(1+2\beta + 2\mathcal{H}A + 2 \frac{\delta K}{a} \frac{\partial a}{\partial K} -\frac{1}{4} \delta K \, {\bf x}^2   \right){\bf \mathrm{d}x}^2 \Bigg]+ ... \, .
\end{eqnarray}
Comparing this to a spatially flat geometry with spatially-homogeneous scalar perturbations in longitudinal gauge,
\begin{equation}
\mathrm{d}s^2 = a^2(\tau) \left[ -(1-2 \hat{\Phi}) \mathrm{d}\tau^2 + (1+2 \hat{\Psi}) \mathrm{d}{\bf x}^2 \right] \, ,
\end{equation}
we see that we must require $\beta' = \delta K = 0$ (i.e. $\beta=\;$constant), and 
\begin{eqnarray}
\hat{\Phi} = -A' - \mathcal{H}A \qquad {\rm and} \qquad
\hat{\Psi} = \beta + \mathcal{H}A \, . 
\end{eqnarray}
The metric in \eqref{metricA} can therefore be thought of as a Robertson-Walker metric with the super-horizon scalar perturbations $\hat{\Phi}$ and $\hat{\Psi}$ given above. These exactly reproduce the expected equations presented in (\ref{large1}) and (\ref{large2}), when taking the $k\rightarrow 0$ limit of the relevant equations from Ref. \cite{Sanghai_2019}.


\section*{References}

\begin{thebibliography}{10}

\bibitem{Ishak_2018}
Mustapha Ishak.
\newblock Testing {G}eneral {R}elativity in cosmology.
\newblock {\em Living Reviews in Relativity}, 22(1), Dec 2018.

\bibitem{GW150914}
B.~P. Abbott et~al.
\newblock Tests of {G}eneral {R}elativity with {GW}150914.
\newblock {\em Phys. Rev. Lett.}, 116:221101, May 2016.

\bibitem{Will_1993}
Clifford~M. Will.
\newblock Theory and {E}xperiment in {G}ravitational {P}hysics.
\newblock Mar 1993.

\bibitem{Poisson_2014}
Eric Poisson and Clifford~M. Will.
\newblock {\em Gravity: Newtonian, Post-Newtonian, Relativistic}.
\newblock Cambridge University Press, 2014.

\bibitem{Will_2014}
Clifford~M. Will.
\newblock The confrontation between {G}eneral {R}elativity and experiment.
\newblock {\em Living Reviews in Relativity}, 17:4, Dec 2014.

\bibitem{Bertotti_2003}
B.~Bertotti, L.~Iess, and P.~Tortora.
\newblock A test of {G}eneral {R}elativity using radio links with the {C}assini
  spacecraft.
\newblock {\em Nature}, 425:374--376, Sep 2003.

\bibitem{Biswas_2008}
Abhijit Biswas and Krishnan R.~S. Mani.
\newblock Relativistic perihelion precession of orbits of {V}enus and the
  {E}arth.
\newblock {\em Central European Journal of Physics}, 6:754--758, May 2008.

\bibitem{Gravity_Probe_B}
C.~W.~F. Everitt et~al.
\newblock Gravity {P}robe {B}: Final results of a space experiment to test
  {G}eneral {R}elativity.
\newblock {\em Phys. Rev. Lett.}, 106:1101, May 2011.

\bibitem{Weisberg_1984}
J.~M. Weisberg and J.~H. Taylor.
\newblock Observations of post-newtonian timing effects in the binary pulsar
  psr 1913+16.
\newblock {\em Phys. Rev. Lett.}, 52:1348--1350, Apr 1984.

\bibitem{Sanghai_2015}
Viraj~A.{\hspace{0.167em}}A. Sanghai and Timothy Clifton.
\newblock Post-{N}ewtonian cosmological modelling.
\newblock {\em Physical Review D}, 91(10):103532, May 2015.

\bibitem{Sanghai_2017}
Viraj A~A Sanghai and Timothy Clifton.
\newblock Parameterized post-{N}ewtonian cosmology.
\newblock {\em Classical and Quantum Gravity}, 34(6):065003, Feb 2017.

\bibitem{Sanghai_2019}
Timothy Clifton and Viraj A.~A. Sanghai.
\newblock Parametrizing theories of gravity on large and small scales in
  cosmology.
\newblock {\em Phys. Rev. Lett.}, 122:011301, Jan 2019.

\bibitem{Hu_2007}
Hu, Wayne and Sawicki, Ignacy.
\newblock A Parameterized Post-Friedmann Framework for Modified Gravity.
\newblock {\em Phys. Rev. D}, 76:104043, Aug 2007. 

\bibitem{Hu_2008}
Wayne Hu.
\newblock Parametrized post-{F}riedmann signatures of acceleration in the CMB.
\newblock {\em Physical Review D}, 77(10):103524, May 2008.

\bibitem{Amin_2008}
Amin, Mustafa A., Wagoner, Robert V. and Blandford, Roger D.
\newblock  A subhorizon framework for probing the relationship between the cosmological matter distribution and metric perturbations.
\newblock {\em Monthly Notices of the Royal Astronomical Society}, 390(1):131--142, Oct 2008.

\bibitem{Skordis_2009}
Constantinos Skordis.
\newblock Consistent cosmological modifications to the {E}instein equations.
\newblock {\em Phys. Rev. D}, 79(12):123527, June 2009. 

\bibitem{Baker_2011}
Baker, Tessa, Ferreira, Pedro G., Skordis, Constantinos and Zuntz, Joe.
\newblock Towards a fully consistent parametrization of modified gravity.
\newblock {\em Phys. Rev. D}, 84(12):124018, Dec 2011.

\bibitem{Baker_2013}
Baker, Tessa, Ferreira, Pedro G. and Skordis, Constantinos.
\newblock The parameterized post-Friedmann framework for theories of modified gravity: Concepts, formalism, and examples.
\newblock {\em Phys. Rev. D}, 87(2):024015, Jan 2013.

\bibitem{Ellis_1999}
George F.~R. {Ellis} and Henk {van Elst}.
\newblock Cosmological models (carg{\`e}se lectures 1998).
\newblock In Marc {Lachi{\`e}ze-Rey}, editor, {\em Theoretical and
  Observational Cosmology}, volume 541 of {\em NATO Advanced Study Institute
  (ASI) Series C}, pages 1--116, January 1999.

\bibitem{Malik_2009}
Karim~A. Malik and David Wands.
\newblock Cosmological perturbations.
\newblock {\em Physics Reports}, 475(1-4):1--51, May 2009.

\bibitem{Kodama_1984}
Hideo Kodama and Misao Sasaki.
\newblock {Cosmological Perturbation Theory}.
\newblock {\em Progress of Theoretical Physics Supplement}, 78:1--166, 01 1984.

\bibitem{Bertschinger_2006}
Edmund Bertschinger.
\newblock On the growth of perturbations as a test of dark energy and gravity.
\newblock {\em The Astrophysical Journal}, 648(2):797--806, sep 2006.

\bibitem{Sanghai_2016}
Viraj A.~A. Sanghai and Timothy Clifton.
\newblock Cosmological backreaction in the presence of radiation and a
  cosmological constant.
\newblock {\em Phys. Rev. D}, 94:023505, Jul 2016.

\bibitem{Goldberg_2017b}
Sophia~R. Goldberg, Christopher~S. Gallagher, and Timothy Clifton.
\newblock Perturbation theory for cosmologies with nonlinear structure.
\newblock {\em Phys. Rev. D}, 96:103508, Nov 2017.

\bibitem{Uzan_2003}
Jean-Philippe Uzan.
\newblock The fundamental constants and their variation: observational and
  theoretical status.
\newblock {\em Rev. Mod. Phys.}, 75:403--455, Apr 2003.

\bibitem{Joudaki_2017}
{Joudaki} et~al.
\newblock {KiDS-450 + 2dFLenS: Cosmological parameter constraints from weak
  gravitational lensing tomography and overlapping redshift-space galaxy
  clustering}.
\newblock {\em Monthly Notices of the Royal Astronomical Society},
  474(4):4894--4924, 10 2017.

\bibitem{Shao_2013}
Lijing {Shao} and Norbert {Wex}.
\newblock {New limits on the violation of local position invariance of
  gravity}.
\newblock {\em Classical and Quantum Gravity}, 30(16):165020, Aug 2013.

\bibitem{Shao_2012}
Lijing {Shao} and Norbert {Wex}.
\newblock {New tests of local Lorentz invariance of gravity with
  small-eccentricity binary pulsars}.
\newblock {\em Classical and Quantum Gravity}, 29:21.5018, October 2012.

\bibitem{Planck_2020}
{Planck Collaboration} et~al.
\newblock Planck 2018 results - vi. cosmological parameters.
\newblock {\em A\&A}, 641, 2020.

\bibitem{Clifton_2020}
Timothy Clifton, Christopher~S. Gallagher, Sophia Goldberg, and Karim~A. Malik.
\newblock Viable gauge choices in cosmologies with nonlinear structures.
\newblock {\em Phys. Rev. D}, 101:063530, Mar 2020.

\bibitem{Thomas_2015}
Daniel~B. Thomas, Marco Bruni, and David Wands.
\newblock {The fully non-linear post-Friedmann frame-dragging vector potential:
  magnitude and time evolution from N-body simulations}.
\newblock {\em Monthly Notices of the Royal Astronomical Society},
  452(2):1727--1742, 07 2015.

\bibitem{Milillo_2015}
Irene Milillo, Daniele Bertacca, Marco Bruni, and Andrea Maselli.
\newblock Missing link: {A} nonlinear post-{F}riedmann framework for small and
  large scales.
\newblock {\em Physical Review D}, 92(2):023519, Jul 2015.

\bibitem{Sampson_2013}
Laura Sampson, Nicol\'as Yunes, and Neil Cornish.
\newblock Rosetta stone for parametrized tests of gravity.
\newblock {\em Phys. Rev. D}, 88:064056, Sep 2013.

\bibitem{Clifton_2021}
Timothy Clifton and Daniel~B. Thomas.
\newblock in prep.

\bibitem{Vainshtein_1972}
A.I. Vainshtein.
\newblock To the problem of nonvanishing gravitation mass.
\newblock {\em Physics Letters B}, 39(3):393--394, 1972.

\bibitem{Khoury_2004}
Justin Khoury and Amanda Weltman.
\newblock Chameleon fields: Awaiting surprises for tests of gravity in space.
\newblock {\em Phys. Rev. Lett.}, 93:171104, Oct 2004.

\bibitem{Khoury_2010}
Kurt Hinterbichler and Justin Khoury.
\newblock Screening long-range forces through local symmetry restoration.
\newblock {\em Phys. Rev. Lett.}, 104:231301, Jun 2010.

\bibitem{Avilez_Lopez_2015}
A.~Avilez-Lopez, A.~Padilla, Paul~M. Saffin, and C.~Skordis.
\newblock The parametrized post-newtonian-vainshteinian formalism.
\newblock {\em Journal of Cosmology and Astroparticle Physics},
  2015(06):044--044, jun 2015.

\bibitem{Clifton_2008}
Timothy Clifton.
\newblock Parametrized post-newtonian limit of fourth-order theories of
  gravity.
\newblock {\em Phys. Rev. D}, 77:024041, Jan 2008.

\bibitem{Clifton_2011}
Timothy Clifton.
\newblock Cosmology without averaging.
\newblock {\em Classical and Quantum Gravity}, 28(16):164011, Aug 2011.

\bibitem{Brizuela_2009}
David Brizuela, José~M. Martín-García, and Guillermo~A. Mena~Marugán.
\newblock x{P}ert: computer algebra for metric perturbation theory.
\newblock {\em General Relativity and Gravitation}, 41(41), Feb 2009.

\bibitem{Pitrou_2013}
Cyril Pitrou, Xavier Roy, and Obinna Umeh.
\newblock x{P}and: an algorithm for perturbing homogeneous cosmologies.
\newblock {\em Classical and Quantum Gravity}, 30(16):165002, Jul 2013.

\bibitem{Hohmann_2021}
Manuel Hohmann.
\newblock {xPPN}: an implementation of the parametrized post-newtonian
  formalism using x{A}ct for mathematica.
\newblock {\em The European Physical Journal C}, 81:504, 2021.

\end{thebibliography}
\end{document}